\documentstyle[12pt]{article}
\def\hot{\hat{\otimes}}
\def\condzu{$z_j=q^{-2(j-1)}z (1\leq j\leq n+1)$,
$u_j=q^{-2(j-1)+1}z (1\leq j\leq n)$}
\def\Blo{B(\Lambda_0)}

\def\blo{b_{\Lambda_0}}

\def\pe{{\cal P}}
\def\la{\lambda}
\def\U{U_q({\hat sl}_2)}
\def\Up{U_q^{'}({\hat sl}_2)}
\def\Upi{U_q^{'}(({\hat sl}_2)_i)}

\def\La{\Lambda}
\def\bla{b_{\Lambda_0}}
\def\blai{b_{\Lambda_1}}

\def\et{\tilde{e}}
\def\ft{\tilde{f}}
\def\ftj{\tilde{f}_j}
\def\fti{\tilde{f}_i}
\def\ftim{\tilde{f}_{1-i}}
\def\eti{\tilde{e}_i}
\def\etim{\tilde{e}_{1-i}}
\def\etj{\tilde{e}_j}
\def\Bi{B_1}
\def\Bs{B_s}
\def\Bsm{B_{s-1}}

\def\Vli{V(\La_i)}
\def\Vhli{\hat{V}(\La_{i})}
\def\Vhlip{\hat{V}(\La_{i+1})}
\def\Vlip{V(\La_{i+1})}
\def\Vljp{V(\La_{j+1})}
\def\Vlj{V(\La_j)}
\def\Viz{(V_1)_z}
\def\Vn{V_n}
\def\Vnp{V_{n+1}}
\def\Vmp{V_{m+1}}
\def\Vm{V_m}

\def\fbj{\fbox{j}}
\def\fbs{\fbox{s}}
\def\fbjm{\fbox{j-1}}
\def\fbjp{\fbox{j+1}}
\def\fbp{\fbox{+}}
\def\fbm{\fbox{-}}
\def\fbo{\fbox{0}}
\def\ra{\longrightarrow}
\def\ot{\otimes}
\def\fbjs{\fbox{j}_{s}}
\def\fbjprs{\fbox{j'}_{s}}
\def\fbjprsm{\fbox{j'}_{{s-1\over2}}}
\def\fbss{\fbox{s}_{{s\over2}}}
\def\fbjsm{\fbox{j}_{s-1}}

\def\fbjps{\fbox{j+1}_{s}}

\def\fbjmsm{\fbox{j-1}_{s-1}}
\def\fbsmsm{\fbox{s-1}_{s-1}}
\def\fbos{\fbox{0}_{{s\over2}}}
\def\fbosm{\fbox{0}_{s-1}}
\def\psit{\tilde{\psi}}
\def\Ub{U'(b_{+})}
\def\qquli{{\bf Q}(q)u_{\Lambda_i}}
\def\uli{u_{\Lambda_i}}
\def\udli{u_{\Lambda_i}^\ast}
\def\udlip{u_{\Lambda_{i+1}}^\ast}
\def\ulim{u_{\Lambda_{1-i}}}

\def\vnj{v^{(n)}_j}
\def\vij{v^{(1)}_j}
\def\vdij{v^{(1)\ast}_j}
\def\vnpj{v^{(n+1)}_j}
\def\vnpk{v^{(n+1)}_k}
\def\vnk{v^{(n)}_k}

\def\vnjp{v^{(n)}_{j+1}}
\def\vnjm{v^{(n)}_{j-1}}
\def\vnnj{v^{(n)}_{n-j}}
\def\vdnj{v^{(n)\ast}_j}
\def\vdnnj{v^{(n)\ast}_{n-j}}
\def\vno{v^{(n)}_0}
\def\uqi{U'_{q}((\hat{sl}_2)_1)}
\def\wj{w_j}

\def\ouunnp{{}^{n}O^{n+1}}
\def\oudnnp{{}^{n}O_{n+1}}
\def\odunnp{{}_{n}O^{n+1}}
\def\phinnp{{}_{V_n}\Phi^{V_{n+1}}}
\def\phinnpad{{}^{V_n}\Phi^{V_{n+1}^{\ast a}}}
\def\phinamdnp{{}^{V_n^{\ast a^{-1}}}\Phi^{V_{n+1}}}
\def\udphinnp{{}^{V_n}\Phi_{V_{n+1}}}
\def\phiunnp{{}^{V_n}\Phi^{V_{n+1}}}

\def\phidusms{{}_{V_{s-1}}\Phi^{V_{s}}}
\def\phiudsms{{}^{V_{s-1}}\Phi_{V_{s}}}

\def\inv{\hbox{inv}}
\def\ep{\epsilon}
\def\H{{\cal H}}
\def\ozu{O({\bf z}|{\bf u})}
\def\otzu{\tilde{O}({\bf z}|{\bf u})}
\def\ojzu{O_j({\bf z}|{\bf u})}
\def\okzu{O_k({\bf z}|{\bf u})}
\def\orzu{O_r({\bf z}|{\bf u})}
\def\oizu{O_1({\bf z}|{\bf u})}
\def\oqtzpup{O(q^{-2}{\bf z}^\prime|q^{-2}{\bf u}^\prime)}
\def\br{\bar{R}}
\def\brc{\check{\bar{R}}}
\def\tr{\tilde{R}}
\def\ttr{\tilde{\tilde{R}}}
\def\oz{O(z)}
\def\bnnj{\left[\begin{array}{c}n\\j\end{array}\right]}
\def\bnjk{\left[\begin{array}{c}j\\k\end{array}\right]}
\def\hom{\hbox{Hom}}
\def\homup{\hbox{Hom}_{\Up}}
\def\id{\hbox{id}}

\def\phiad{\Phi^{V^{\ast a}}}

\def\psiamd{\Psi^{V^{\ast a^{-1}}}}
\def\otzzuu{\tilde{O}({\bf z}^\prime,{\bf z}|{\bf u}^\prime,{\bf u})}

\title{
Fusion of the q-Vertex Operators and its
Application\\
to Solvable Vertex Models}

\author{
Atsushi Nakayashiki\\
Graduate School of Mathematics, \\
Kyushu University,\\
Ropponmatsu 4-2-1, Fukuoka 810\\
e-mail:6vertex@math.kyushu-u.ac.jp}

\date{August, 1994}
\begin{document}
\maketitle
\begin{abstract}
We diagonalize the transfer matrix of the inhomogeneous vertex
models of the 6-vertex type in the anti-ferroelectric regime
intoducing new types of q-vertex operators.
The special cases of those models were used to diagonalize
the s-d exchange model\cite{W,A,FW1}. New vertex operators
are constructed from the level one vertex operators
by the fusion procedure and have the description by bosons.
In order to clarify the particle structure we estabish
new isomorphisms of crystals. The results are very simple
and figure out representation theoretically
the ground state degenerations.
\end{abstract}

\section{Introduction}
In \cite{DFJMN} the anti-ferroelectric XXZ hamiltonian,
or equivalently,
the transfer matrix of the 6-vertex model has been diagonalized
directly in the thermodynamic limit based on the quantum
affine symmetry. The method is powerful enough, on the one hand,
to give the integral formulas for correlation functions and
form factors, on the other hand, to determine the physical space
as a representation of a quantum affine algebra.

Similar approach is possible for several two dimensional
lattice models such as ABF model\cite{JMO,FJMMN}.
Among them a direct generalization of the 6-vertex model
is the vertex models associated with the perfect representations
of any level\cite{KKMMNN1,KKMMNN2}.
Although there are technical problems of bosonozation
in the case of higher levels, at least the strategy is clear
and everything we need is in our hands.

In this paper I want to add one more class of vertex models
which can be solved by a similar method and are not contained
in the class of directly generalized models said above.
The vertex models which we study here is the inhomogeneous
vertex models of 6-vertex type with the inhomogeneities
being in the spins.
Namely, on the infinite regular square lattice,
with each horizontal and vertical lines except a finite number
of vertical lines $l_1,\cdots,l_n$, we associate the
vector space ${\bf C}^2$. With $l_1,\cdots,l_n$ we associate
${\bf C}^{s_1+1},\cdots,{\bf C}^{s_n+1}$ for arbitrary non-negative
integers $s_1,\cdots,s_n$. To each vertex the Boltzmann weight
is defined by the corresponding trigonometric $R$-matrix
acting on ${\bf C}^2\ot{\bf C}^2$ or ${\bf C}^2\ot{\bf C}^{s_j}$.
The rational limits of those models with $n=1$ had been used
to diagonalize the s-d exchange models (Kondo problem)\cite{A,W,FW1,TW}.

The central object in the symmetry approach is the
q-vertex operator which was introduced by
Frenkel-Reshetikhin\cite{FR}.
In the case of the 6-vertex model the q-vertex operator
makes it possible to identify the infinite tensor product
$({\bf C}^2)^{\ot {\bf Z}_{\geq1}}$ with the irreducible
representation $\Vli$ of $\U$.
Using this identification, the transfer matrix,
the creation-annihilation operators, correlation functions
and form factors are all described in terms of q-vertex
operators.

Similarly, in our case, everything is described by q-vertex
operators. But here appeares a new phenomenon, the degeneration
of the ground states. To take this effect into consideration is
crucial in the theory. To treate this situation correctly
what we must do is to introduce new kinds of q-vertex
operators. Those new operators can be considered as a mixture
of type I and type II vertex operators in the terminology
of \cite{DFJMN}. Naturally they can be obtained by a fusion
procedure from level one vertex operators. In particular
new operators have the description by free fields.
Hence physical quantities of our models can be written down in the
form of integral formulas. We study these formulas in the
next paper.

Let us describe our story more precicely.
The total quantum space which is acted by the transfer matrix is
\begin{eqnarray}
&&
\oplus_{i,j=0,1}\Vli\ot V_{s_n}\ot\cdots\ot V_{s_1}\ot\Vlj^{\ast a},
\label{spacei}
\end{eqnarray}
where $V_s\simeq {\bf C}^{s+1}$ and considered as the representation
of $\Up$.
In order to give the description of the correlation function
or form factors we must know the structure of eigenstates of the
transfer matrix.
The insight comes, as in the case of the XXZ-model\cite{DFJMN,IIJMNT},
from the decomposition of crystals
\[
B(\La_i)\otimes B_{s_1}\otimes\cdots
\otimes B_{s_k}\otimes B(\La_j)^\ast.
\]
The result is surprisingly simple (see Corollary \ref{ltstr}).
Consequently we find that the physical space of
our models can be written as
\begin{eqnarray}
&&
{\bf C}^{s_n}\ot\cdots\ot {\bf C}^{s_1}
\ot\Big[\oplus_{m=0}^\infty
\int_{|z_1|=1}\cdots\int_{|z_m|=1}({\bf C}^2)^{\ot m}\Big]_{sym},
\nonumber
\end{eqnarray}
where sym is some symmetrization. In this tensor product the last
term which is described by a bracket is the physical space of
the XXZ model. On the other hand former tensor component
describes the ground state degeneration. In the case $n=1$
the dimension of the degeneracy of the ground states coinsides
with the results of Fateev-Wiegman\cite{FW1} in the rational limits.
This picture of the structure of the space of states suggests that
it is natural to consider the space
\begin{eqnarray}
&&
\oplus_{i,j=0,1}
V_{s_n-1}\ot\cdots\ot V_{s_1-1}\ot\Vli\ot\Vlj^{\ast a}.
\label{spaceii}
\end{eqnarray}
The relation between two spaces \ref{spacei} and \ref{spaceii}
is given by the vertex operators
\begin{eqnarray}
&&
\phidusms(z):(V_{s-1})_z\ot\Vli\ra\Vlip\ot(V_s)_z,
\label{voi}
\\
&&
\phiudsms(z):\Vli\ot(V_s)_z\ra(V_{s-1})_z\ot\Vlip,
\label{voii}
\end{eqnarray}
which are parts of the newly introduced operators.
On the space \ref{spaceii} descriptions of the
model and physical quantities take very simple forms.
For example the transfer matrix is, up to a scalar multiple, equal
to $1\ot T_{XXZ}(z)$, where $T_{XXZ}(z)$ is the transfer matrix
of the 6-vertex model.
The peculiarity of our model comes from the definition
of the local operators which are defined using vertex operators
\ref{voi} and \ref{voii}.

The present paper is organized in the following manner.
In section 2 we review necessary preliminaries and notations.
In section 3 we establish a new type of isomorphisms
of crystals which is considered to be a generalization
of the path realization of the crystals with highest weights.
Applying this isomorphism we determine the decomposition
of the crystal mentioned above.
In section 4 we introduce a new vertex operators and prove their
existence. The existence theorem explains why the spectral parameters
in the right hand side and in the left hand side
of \ref{voi}, \ref{voii} must coinside.
This fact is important to treat the ground state degeneration.
The fusion construction of the representations and R-matrices
are briefly reviewed in section 5.
In section 6 the fusion procedure is carried out for level one
vertex operators and construct new vertex operators.
The well definedness of the fusion proceadure is the main result
here.
We determine the commutation relations of newly defined vertex
operators using the fusion construction in section 7.
In section 8 we propose the mathematical settings for our models.
In appendix 1 the integral formulas for the highest-highest
matrix element of the composition of type I and type II
vertex operators are given. In appendix 2 ,3 the description
of level one vertex operators in terms of bosons and their
OPEs are given. These are used to derive the integral formulas
in appendix 1.

\section{Notations and preliminaries}
\subsection{Definition of quantized envelopping algebra}
Let us recall the definition of $\U$ and fix several notations
related to it.
Let $P={\bf Z}\La_0\oplus {\bf Z}\La_1\oplus {\bf Z}\delta$,
$P^\ast={\bf Z}h_0\oplus {\bf Z}h_1\oplus {\bf Z}d$
be the weight and the dual weight lattice of $\hat{sl}_2$
with the pairing
$<\La_i,h_j>=\delta_{ij}$, $<\La_i,d>=0$, $<\delta,h_i>=0$,
$<\delta,d>=1$.
Set $\alpha_1=2\La_1-2\La_0$, $\alpha_0=\delta-\alpha_1$,
$\rho=\La_0+\La_1$.
The symmetric bilinear form on $P$
normalized as $(\alpha_i,\alpha_i)=2$ is given by
$(\La_i,\La_j)={\delta_{1i}\delta_{1j}\over2}$,
$(\La_i,\delta)=1$, $(\delta,\delta)=0$.
Through $(,)$ we consider $P^\ast$ as a subset of $P$
so that $2\rho=h_1+4d$.
Let us set $F={\bf Q}(q)$ with $q$ being the complex number
transcendent over the rational number field ${\bf Q}$.
In section 8, we assume that the $q$ is real and $-1<q<0$.

The algebra $\U$ is the $F$-algebra generated by
$e_i,f_i,(i=0,1),q^{h}(h\in P^\ast)$ with the defining relations
\begin{eqnarray}
&&
q^0=1,
\quad
q^{h_1}q^{h_2}=q^{h_1+h_2},
\quad
q^he_iq^{-h}=q^{<h,\alpha_i>}e_i,
\quad
q^hf_iq^{-h}=q^{-<h,\alpha_i>}f_i,
\nonumber
\\
&&
[e_i,f_j]={t_i-t_i^{-1}\over q-q^{-1}},
\quad
\sum_{m=0}^3(-1)^mx_i^{(m)}x_jx_i^{(3-m)}=0 \quad(i\neq j)
\hbox{ for $x=e,f$}
\nonumber
\end{eqnarray}
where we set $t_i=q^{h_i}$ and
\begin{eqnarray}
&&
x_i^{(m)}={x_i^m\over [m]!}\quad(x=e,f),
\quad
[m]={q^m-q^{-m}\over q-q^{-1}},
\quad
[m]!=\prod_{k=1}^m[k],
\quad
\bnnj={[n]!\over [j]![n-j]!}.
\nonumber
\end{eqnarray}
We denote by $U'=\Up$ the subalgebra of $\U$ generated
by $e_i$, $f_i$, $t_i$ $(i=0,1)$.

\subsection{Hopf algebra structure}
In this paper we use the following coproduct and the anti-pode,
\begin{eqnarray}
&&
\Delta(e_i)=e_i\ot 1+t_i\ot e_i,
\quad
\Delta(f_i)=f_i\ot t_i^{-1}+1\ot f_i,
\quad
\Delta(q^{h})=q^h\ot q^h,
\nonumber
\\
&&
a(e_i)=-t_i^{-1}e_i,
\quad
a(f_i)=-f_it_i,
\quad
a(q^h)=q^{-h}.
\nonumber
\end{eqnarray}

\subsection{Finite dimensional module}
The $\Up$ module
$(\Vn)_z=\oplus_{j=0}^nF[z,z^{-1}]\vnj$ is defined as
\begin{eqnarray}
&&
f_1\vnj=[n-j]\vnjp,
\quad
e_1\vnj=[j]\vnjm,
\quad
t_1\vnj=q^{n-2j}\vnjp,
\nonumber
\\
&&
f_0=z^{-1}e_1,
\quad
e_0=zf_1,
\quad
t_0=t_1^{-1},
\nonumber
\end{eqnarray}
where $z$ is a complex number.
In particular $\vnj=\bnnj^{-1}{\vno\over[j]!}$.
In the following sections, for the sake of simplicity,
we only write $F$ instead of $F[z,z^{-1}]$ as far as
no confusion occurs in every situation.

\subsection{Dual module}
For a left $\Up$-module $M$, we define the left module $M^{\ast a^{\pm1}}$
by
\begin{eqnarray}
&&
M^{\ast a^{\pm1}}=\hom(M,F)\hbox{ as a linear space,}
\nonumber
\\
&&
<xw,v>=<w,a^{\pm1}(x)v>
\hbox{ for $w\in M^{\ast a^{\pm1}}$, $v\in M$ and $x\in\Up$.}
\nonumber
\end{eqnarray}
Here the linear dual of an integrable module with finite dimensional
weight spaces should be considered to be the restricted dual.
By definition $M$, $M^{\ast a\ast a^{-1}}$ and $M^{\ast a^{-1}\ast a}$
are canonically isomorphic.
For these dual modules the following properties hold,
\begin{eqnarray}
&&
\homup(M_1\ot M_2,M_3)\simeq\homup(M_1,M_3\ot M_2^{\ast a}),
\nonumber
\\
&&
\homup(M_1\ot M_2,M_3)\simeq\homup(M_2,M_1^{\ast a^{-1}}\ot M_3),
\nonumber
\end{eqnarray}
where $\homup(M_1,M_2)$ is the vector space of $\Up$ linear
homomorphisms.
Let $\{\vdnj\}$ be the dual base of
$\{\vnj\}$, $<\vdnj,\vnk>=\delta_{jk}$.
Then the following isomorphisms hold,
\begin{eqnarray}
(\Vn)_{q^{\mp2}z}
&\simeq&
(\Vn)_z^{\ast a^{\pm1}}
\label{selfdual}
\\
\vnj
&\mapsto&
(-)^jq^{-j(n-j\mp1)}\bnnj^{-1}\vdnnj.
\nonumber
\\
(-)^{n-j}q^{(n-j)(j\mp1)}\bnnj\vnnj
&\leftarrow&
\vdnj.
\nonumber
\end{eqnarray}

\subsection{Level one vertex opoerators}
Let $\Vli$ be the irreducible highest weight $\Up$-module with
highest weight $\La_i (i=0,1)$, $\Vhli$ its weight completioin
$\Vhli=\prod_{\nu\in P}\Vli_{\nu}$,
$\Vli_{\nu}=\{v\in\Vli|q^hv=q^{<h,\nu>}v \hbox{ for }$ $h\in P\}$
and $\udli$ the highest weight vector of the right module
$\Vli^\ast$ such that $<\udli,\uli>=1$.
Let us denote $\Phi(z)$, $\Psi(z)$, $\Phi_V(z)$ and $\Psi_V(z)$
the $\Up$ intertwiners
\begin{eqnarray}
&&
\Phi(z): \Vli\ra\Vlip\hot \Viz,
\nonumber
\\
&&
\Psi(z): \Vli\ra\Viz\hot\Vlip,
\nonumber
\\
&&
\Phi_V(z): \Vli\ot\Viz\ra\Vhlip,
\nonumber
\\
&&
\Psi_V(z): \Viz\ot\Vli\ra\Vhlip,
\nonumber
\end{eqnarray}
normalized as
\begin{eqnarray}
&&
<\udlip,\Phi(z)\uli>=
<\udlip,\Psi(z)\uli>=v^{(1)}_{1-i},
\nonumber
\\
&&
<\udlip,\Phi_V(z)(\uli\ot v^{(1)}_{j})>=
<\udlip,\Psi_V(z)(\uli\ot v^{(1)}_{j})>=\delta_{i,j}.
\nonumber
\end{eqnarray}
Here and after, in general, we set
$\Vli\hot(V_n)_z=
\big(\prod_{\nu\in P}F[z,z^{-1}]\ot\Vli_{\nu}\big)
\ot_{F[z,z^{-1}]}(V_n)_z$.
In fact the images of $\Phi(z)$ and $\Psi(z)$ belong to smaller
spaces\cite{DFJMN}.

The components of those operators are defined by
\begin{eqnarray}
&&
\Phi_j(z)=<v^{(1)\ast}_j,\Phi(z)>,
\quad
\Psi_j(z)=<v^{(1)\ast}_j,\Psi(z)>.
\nonumber
\end{eqnarray}

We shall also introduce the notations
$\phiad(z)$ and $\psiamd(z)$ which are intertwiners
\begin{eqnarray}
&&
\phiad(z): \Vli\ra\Vlip\hot \Viz^{\ast a},
\nonumber
\\
&&
\psiamd(z): \Vli\ra\Viz^{\ast a^{-1}}\hot\Vlip,
\nonumber
\end{eqnarray}
defined by
\begin{eqnarray}
&&
<\vij,\phiad(z)u>=\Phi_V(z)(u\ot \vij),
\quad
<\vij,\psiamd(z)u>=\Psi_V(z)(\vij\ot u).
\nonumber
\end{eqnarray}
Those operators are related by
\begin{eqnarray}
&&
\phiad(z)=(-1)^{1-i}q^{-1+i}\Phi(q^{-2}z),
\quad
\psiamd(z)=(-1)^{1-i}q^{1-i}\Psi(q^2z),
\nonumber
\end{eqnarray}
under the isomorphism \ref{selfdual}.

The commutation relations of those vertex operators are,
on $\Vli$,
\begin{eqnarray}
&&
-({z_1\over z_2})^ir({z_1\over z_2})\brc({z_1\over z_2})
\Phi(z_1)\Phi(z_2)=\Phi(z_2)\Phi(z_1),
\nonumber
\\
&&
({z_1\over z_2})^{1-i}r({z_1\over z_2})
\brc({z_1\over z_2})
\Psi(z_2)\Psi(z_1)=\Psi(z_1)\Psi(z_2),
\nonumber
\\
&&
({z_1\over z_2})^{-1+i}
{\theta_{q^4}({qz_1\over z_2})\over\theta_{q^4}({qz_2\over z_1})}
\Psi(z_1)\Phi(z_2)=\Phi(z_2)\Psi(z_1).
\nonumber
\end{eqnarray}
We shall rewrite the first and second relations
 for the sake of later use as
\begin{eqnarray}
&&
q({z_1\over z_2})^ir({q^2z_1\over z_2})\brc({q^2z_1\over z_2})
\Phi(z_1)\phiad(z_2)=\phiad(z_2)\Phi(z_1),
\nonumber
\\
&&
-q^{-1}({z_1\over z_2})^{1-i}r({z_1\over q^2z_2})
\brc({z_1\over q^2z_2})
\psiamd(z_2)\Psi(z_1)=\Psi(z_1)\psiamd(z_2).
\nonumber
\end{eqnarray}
Here $\brc(z)=P\br(z)$, $P(u\ot v)=v\ot u$ and
\begin{eqnarray}
&&
\br(z)(\vij\ot\vij)=\vij\ot\vij \hbox{ for $j=0,1$},
\nonumber
\\
&&
\br(z)(v^{(1)}_0\ot v^{(1)}_1)=
bv^{(1)}_0\ot v^{(1)}_1+cv^{(1)}_1\ot v^{(1)}_0,
\nonumber
\\
&&
\br(z)(v^{(1)}_1\ot v^{(1)}_0)=
c^\prime v^{(1)}_0\ot v^{(1)}_1+bv^{(1)}_1\ot v^{(1)}_0,
\nonumber
\\
&&
b={1-z\over 1-q^2z}q,
\quad
c={1-q^2\over 1-q^2z}z,
\quad
c^\prime={1-q^2\over 1-q^2z},
\quad
r(z)=
{(z^{-1})_\infty(q^2z)_\infty
\over
(z)_\infty(q^2z^{-1})_\infty},
\nonumber
\end{eqnarray}
where $(z)_\infty=\prod_{j=0}^\infty(1-zq^{4j})$.

Let us describe the inversion relations for vertex operators.
Let us denote by $P^1_F$ the dual pairing map
$\Viz^{\ast a}\ot \Viz\ra F$ which is in fact a $\Up$ linear.
Then we have
\begin{eqnarray}
&&
P^1_F\phiad(z)\Phi(z)=g^{-1}\id_{\Vli},
\label{invi}
\\
&&
Res_{z_1=z_2}\Psi(z_2)\psiamd(z_1)=
z_2gw\ot\id_{\Vli},
\label{invii}
\end{eqnarray}
where $g={(q^2)_\infty\over (q^4)_\infty}$ and
\begin{eqnarray}
&&
w=\sum_{j=0}^1v^{(1)\ast}_j\ot v^{(1)}_j=
v^{(1)}_0\ot v^{(1)}_1-qv^{(1)}_1\ot v^{(1)}_0
\nonumber
\end{eqnarray}
through the isomorphism \ref{selfdual}.
Note that \ref{invi} and \ref{invii} are equivalent, respectively, to
\begin{eqnarray}
&&
\Phi_V(z)\Phi(z)=g^{-1}\id_{\Vli}.
\nonumber
\\
&&
Res_{z_1=q^2z_2}\Psi(z_2)\Psi(z_1)=(-1)^{1-i}q^{i+1}z_2
gw\ot \id_{\Vli}.
\label{typeii}
\end{eqnarray}

\subsection{Crystal}
We shall review here the definitions and fundamental properties
of crystals which we need in the subsequent sections.
The details and generalities
of this section can be found in \cite{KKMMNN1,K2,K3}.

\newtheorem{defi}{Definition}
\begin{defi}
An affine crystal $B$ is a set $B$ with the weight
decomposition $B=\sqcup_{\la\in P}B_\la$ and with
the maps
\begin{eqnarray}
&&
\eti,\fti: B\sqcup \{0\}\ra B\sqcup \{0\}
\nonumber
\end{eqnarray}
satisfying the following axioms:
\begin{description}
\item[(1)] $\eti B_\la\subset B_{\alpha_i+\la}\sqcup \{0\},
\quad
\fti B_\la\subset B_{-\alpha_i+\la}\sqcup \{0\}$ for
non-empty $B_\la$,
\item[(2)]
$\eti0=\fti0=0$,
\item[(3)]
for any $b$ and $i$, there exists $n$ such that
$\eti^nb=\fti^nb=0$,
\item[(4)]
for $b_1,b_2\in B$, $b_2=\fti b_1$ if and only if $b_1=\eti b_2$,
\item[(5)]
if we set
\begin{eqnarray}
&&
\varphi_i(b)=max\{n|\fti^nb\in B\},
\quad
\varepsilon_i(b)=max\{n|\eti^nb\in B\},
\nonumber
\end{eqnarray}
then $\varphi_i(b)-\varepsilon_i(b)=<h_i,\la>$ for
$b\in B_{\la}$ and $i$.
\end{description}
\end{defi}

Let us set $P_{cl}=P/{\bf Z}\delta$ and $cl$ the projection
$P\ra P_{cl}$. Then a classical
crystal is defined using $P_{cl}$ in stead of
$P$ in the definition of an affine crystal.
In this paper crystal means affine or classical crystal.
A slightly general definition of the concept of crystal
is introduced in \cite{K2,K3}.

A crystal has the structure of colored oriented graph by
\begin{eqnarray}
&&
b_1\stackrel{i}{\ra}b_2
\hbox{ if and only if }
b_2=\fti b_1.
\nonumber
\end{eqnarray}

A morphism $\psi: B^1\ra B^2$ of the crystals is a map
$B^1\sqcup \{0\}\ra B^2\sqcup \{0\}$ which commutes with the actions
of $\eti$ and $\fti$ and satisfies $\psi(0)=0$.
A morphism of crystals is called isomorphism (injective) if the accociated
map is bijective (injective). A crystal $B_1$ is called a subcrystal
of $B^2$ if there is an injective morphism of crystals $B^1\ra B^2$.

For a crystal $B$ and a subset $I\subset \{0,1\}$, the $I$-crystal
$B$ is the set $B$ with the same weight decomposition
and with the maps $\etj$, $\ftj$ $(j\in I)$ which is a part of the
maps of the original crystal $B$.

For two crystals $B_1$, $B_2$ we can define the tensor product
in the following manner.

\begin{defi}
\begin{description}
\item[(1)]
As a set $B^1\ot B^2=\sqcup_{\la\in P}(B^1\ot B^2)_\la$,
$(B^1\ot B^2)_\la=\sqcup_{\mu+\nu=\la}B^1_\mu\times B^2_\nu$.
 We denote $(b_1,b_2)$
by $b_1\ot b_2$.
\item[(2)]
The actions of $\eti$ and $\fti$ is defined as
\begin{eqnarray}
&&
\fti(b_1\ot b_2)=\left\{
\begin{array}{ll}
\fti b_1\ot b_2
&
\varphi_i(b_1)>\varepsilon_i(b_2)
\\
b_1\ot \fti b_2
&
\varphi_i(b_1)\leq\varepsilon_i(b_2)
\end{array}
\right.
\nonumber
\\
&&
\eti(b_1\ot b_2)=\left\{
\begin{array}{ll}
\eti b_1\ot b_2
&
\varphi_i(b_1)\geq \varepsilon_i(b_2)
\\
b_1\ot \eti b_2
&
\varphi_i(b_1)<\varepsilon_i(b_2)
\end{array}
\right.
\nonumber
\end{eqnarray}
\end{description}
\end{defi}

Among the crystals we need, in this paper, three kinds of crystals.
The first one is the crystal $B_s$ associated with the crystal base
of the representation $(V_s)_1$.
More explicitly $B_s$ can be described as

\begin{defi}
\begin{description}
\item[(1)] $B_s=\{\fbj|0\leq j\leq s\}$ as a set.
\item[(2)] $\tilde{f_1}\fbj=\fbjp\quad (0\leq j\leq s-1),
\quad
\tilde{f_0}\fbj=\fbjm\quad (1\leq j\leq s),
\quad
\fti\fbj=0\hbox{ otherwise}.$
\end{description}
\end{defi}

In the following we often use the notations $B_1=\{\fbp,\fbm\}$
by the correspondence $\fbp\leftrightarrow\fbox{0}$,
$\fbm\leftrightarrow\fbox{1}$ and identify $\pm$ with $\pm1$.

The second one is the affine crystal $Aff(B_s)$ which is called
the affinization of $B_s$. $Aff(B_s)$ is defined as
\begin{eqnarray}
&&
Aff(B_s)=\sqcup_{\la\in P}Aff(B_s)_\la,
\quad
Aff(B_s)_\la=(B_s)_{cl(\la)}.
\nonumber
\end{eqnarray}
The actions of $\eti,\fti$ are specified by the commutative diagrams
\begin{equation}
\begin{array}{ccccccc}
Aff(B_s)_\la
&
\stackrel{\eti}{\ra}
&
Aff(B_s)_{\la+\alpha_i}\sqcup\{0\}
&
{}
&
Aff(B_s)_\la
&
\stackrel{\fti}{\ra}
&
Aff(B_s)_{\la-\alpha_i}\sqcup\{0\}
\\
\parallel
&
{}
&
\parallel
&
\hbox{and}
&
\parallel
&
{}
&
\parallel
\\
(B_s)_{cl(\la)}
&
\stackrel{\eti}{\ra}
&
(B_s)_{cl(\la+\alpha_i)}\sqcup\{0\}
&
{}
&
(B_s)_{cl(\la)}
&
\stackrel{\fti}{\ra}
&
(B_s)_{cl(\la-\alpha_i)}\sqcup\{0\}.
\end{array}
\nonumber
\end{equation}
For example the graph of $Aff(B_1)$ is
\begin{eqnarray}
&&
\cdots
\stackrel{1}{\ra}
\circ\stackrel{0}{\ra}
\circ\stackrel{1}{\ra}
\circ\stackrel{0}{\ra}
\circ\stackrel{1}{\ra}
\circ\stackrel{0}{\ra}
\cdots.
\nonumber
\end{eqnarray}

Third one is the crystal $B(\La_i)$ associated
with the crystal base of the representation $V(\La_i)$.
By now it is well known that $B(\La_i)$ is described
in terms of the set of paths\cite{KKMMNN1,JMMO}. Let us define
the space of paths $\pe(\La_i)$ as
\begin{eqnarray}
&&
\pe(\La_i)=
\{
(p(j))_{j=1}^\infty|
p(j)\in B_1, p(k)=(-1)^{i+k}\hbox{ for $k>>0$}
\}
\nonumber
\end{eqnarray}
$\pe(\La_i)$ has the structure of an affine crystal
by\cite{JMMO}

\newtheorem{theor}{Theorem}

\begin{theor}\label{fisom}
\begin{description}
\item[(1)] There is an isomorphism of classical crystals,
\begin{equation}
B(\La_i)\simeq B(\La_{1-i})\ot B_1.\label{iteration}
\end{equation}
\item[(2)] The isomorphism \ref{iteration} induces the bijective map
$B(\La_i)\simeq \pe(\La_i)$.
\end{description}
\end{theor}

The weight of a path through the above bijection
can explicitly be written in terms of the energy function \cite{KKMMNN1}.

For an affine crystal $B$ we define the dual crystal $B^\ast$
of $B$ as

\begin{defi}
\begin{description}
\item[(1)]
$B^\ast=\{b^{\vee}|b\in B\}=\sqcup_{\la\in P}B_{-\la},
\quad
B_{-\la}=\{b^{\vee}|b\in B_\la\}$,
\item[(2)]
$\eti b^{\vee}=(\fti b)^{\vee},
\quad
\fti b^{\vee}=(\eti b)^{\vee},
\quad
0^{\vee}=0$.
\end{description}
\end{defi}

The map $(b_1\ot b_2)^\vee\mapsto b_2^\vee\ot b_1^\vee$
gives the isomorphism
\begin{eqnarray}
&&
(B_1\ot B_2)^\ast\simeq
B_2^\ast \ot B_1^\ast.
\nonumber
\end{eqnarray}
Then we have the description of $B(\La_i)^\ast$ in terms
of paths,
\begin{eqnarray}
&&
B(\La_i)^\ast=
\{
(p(j))^{0}_{j=-\infty}|
p(j)\in B_1, p(k)=(-1)^{i+k}\hbox{ for $k<<0$}
\},
\nonumber
\\
&&
B_1\ot B(\La_i)^\ast\simeq B(\La_{i+1})^\ast,
\quad
b\ot(p(j))^0_{j=-\infty}
\mapsto
(p^\prime(j))^0_{j=-\infty},
\nonumber
\end{eqnarray}
where $p^\prime(0)=b$, $p^\prime(j)=p(j+1) (j\leq -1)$.

\subsection{The morphism of crystals induced from the Dynkin diagram
automorphism}\label{Dynkin}
Let $\iota$ be the isomorphism of the ${\bf Z}$ module $P_{cl}$
defined by $\iota(\La_i)=\La_{1-i} (i=0,1)$.
For a classical crystal $B$, we define the classical crystal
$\iota^\ast B$ by
\begin{eqnarray}
&&
\iota^\ast B=\sqcup_{\la\in P_{cl}}(\iota^\ast B)_\la,
\quad
(\iota^\ast B)_\la=\{\iota^\ast(b)|b\in B_{\iota(\la)}\},
\quad,
\iota(0)=0,
\label{DynkinI}
\\
&&
\fti\iota^\ast(b)=\iota^\ast(\ftim b),
\quad
\eti\iota^\ast(b)=\iota^\ast(\etim b).
\label{DynkinII}
\end{eqnarray}
It is easy to prove that \ref{DynkinI}, \ref{DynkinII}
actually defines a classical crystal.
For this crystal the following properties hold.

\newtheorem{prop}{Proposition}
\begin{prop}
\begin{description}
\item[(1)] $\iota^\ast B(\La_i)\simeq B(\La_{1-i})$.
\item[(2)] $\iota^\ast B_s\simeq B_s$ by $\fbj\mapsto \fbox{s-j}$.
\item[(3)] For crystals $B^1$, $B^2$,
$B^1\simeq B^2$ if and only if $\iota^\ast B^1\simeq \iota^\ast B^2$.
\item[(4)] For crystals $B^1$, $B^2$,
$\iota^\ast(B^1\ot B^2)\simeq \iota^\ast B^1 \ot \iota^\ast B^2$,
by $\iota^\ast(b_1\ot b_2)\mapsto \iota^\ast(b_1)\ot \iota^\ast(b_2)$.
\end{description}
\end{prop}

The properties (2)-(4) can directly be checked using definitions.
Let us prove (1).
{}From \ref{iteration}, (2) and (4) above, we have
\begin{eqnarray}
&&
\iota^\ast B(\La_i)\simeq \iota^\ast B(\La_{1-i})\ot B_1,
\quad
\iota^\ast(\uli)\mapsto\iota^\ast(\ulim)\ot\fbox{$(-1)^i$}.
\nonumber
\end{eqnarray}
Using the fact that any element of $\iota^\ast B(\La_i)$
can be written as
$\ft_{j_1}\cdots\ft_{j_n}\iota^\ast(\uli)$
for some $n$ and $(j_1,\cdots,j_n)\in \{0,1\}^n$,
we have $\iota^\ast B(\La_i)\simeq\pe(\La_{1-i})$
as a classical crystal.
$\Box$

\section{Isomorphisms of crystals}\label{crystal}
The structure of the space of
the eigenvectors of the XXZ hamiltonian is,
in the low temprature limit, described
by the decomposition of the crystals of
$B(\La_i)\otimes B(\La_j)^\ast$ \cite{DFJMN}. In this section we shall
prove a new type of isomorphisms of crystals
which generalize Theorem \ref{fisom} (1) and
give a predicted form of the structure of the space
of eigenvectors of our transfer matrix in the low temperature
limit.

Our task is to decompose
the crystals of the form
\[
B(\La_i)\otimes B_{s_1}\otimes\cdots
\otimes B_{s_k}\otimes B(\La_j)^\ast.
\]

The main results in this section are

\begin{theor}\label{mainth}
There is an isomorphism of classical crystals,
\[
B(\La_i)\otimes B_s\simeq
B_{s-1}\otimes B(\La_{1-i}),
\]
for $s=1,2,3,\cdots$.
\end{theor}

\newtheorem{cor}{Corollary}\label{lowtemp}

\begin{cor}\label{ltstr}
For $j=0,1$, we have the isomorphism of classical
crystals,
\begin{eqnarray}
&&\coprod_{i=0,1}
B(\La_i)\otimes B_{s_1}\otimes\cdots
\otimes B_{s_k}\otimes B(\La_j)^\ast
\simeq
\nonumber
\\
&&
B_{s_1-1}\otimes\cdots
\otimes B_{s_k-1}\otimes
\coprod_{n=0}^{\infty}
\hbox{Aff}(B_1)^{\otimes n}/{S_n}.
\nonumber
\end{eqnarray}
\end{cor}

Here the action of the symmetric group $S_n$
on $\hbox{Aff}(B_1)^{\otimes n}$ is
not the usual one but that defined in \cite{DFJMN}.

The isomorphisms of Theorem \ref{mainth} includes \ref{iteration}
as a special case $s=1$. But the proof of Theorem \ref{mainth}
uses the isomorphism \ref{iteration}.

It is sufficient to prove the theorem for $i=0$.
Since $i=1$ case is obtained by applying the map $\iota$
in subsection \ref{Dynkin}.

Let us define the map
\[
\psi:
B_{s-1}\otimes B(\La_0)
\ra
B(\La_1)\otimes B_{s},
\]
first and after that prove that it is well defined and commutes
with the actions of $\eti$ and $\ftj$.
In order to define the map $\psi$ we need to
describe some isomorphisms.

\newtheorem{lemma}{Lemma}

\begin{lemma}\label{map}
There is an isomorphism of $\{0,1\}$ crystals,
\[
\psi_1:\Bs\otimes\Bi
\simeq
\Bi\ot\Bs.
\]
The isomorphism is given explicitly by
\begin{eqnarray}
&&
\fbjp\ot\fbp\ra\fbm\ot\fbj
\hbox{ for $0\leq j\leq s-1$},
\nonumber
\\
&&
\fbo\ot\fbp\ra
\fbp\ot\fbo,
\nonumber
\\
&&
\fbj\ot\fbm\ra
\fbp\ot\fbjp
\hbox{ for $0\leq j\leq s-1$},
\nonumber
\\
&&
\fbs\ot\fbm\ra
\fbm\ot\fbs.
\nonumber
\end{eqnarray}
\end{lemma}

Using the map $\psi_1$ let us define the isomorphism
\[
\psi_n:\Bs\ot\Bi^{\ot n}
\simeq
\Bi^{\ot n}\ot\Bs
\]
by
\[
\psi_n=(1_{n-1}\ot\psi_1)\cdots
(1_1\ot\psi_1\ot1_{n-2})
(\psi_1\ot1_{n-1}),
\]
where $1_j$ is the identity map of $\Bi^{\ot j}$.

Now let us define the map $\psi$ in the following manner.
Take any $\fbj_s\ot b\in \Bsm\ot \Blo$. For $b$ there is
$n\in {\bf Z}_{\geq1}$ which satisfies
\begin{eqnarray}
&&
b=(b_k)_{k=1}^{\infty},
\quad
b_k=(-1)^k
\hbox{ for $k\geq 2n$}.
\label{n}
\end{eqnarray}
Taking any such $n$ and set
\[
\psi(\fbjsm\ot b)=
\blo\ot
\psi_{2n-1}(\fbjs\ot b_{2n-1}\ot\cdots\ot b_1)
\]
throught the isomorphism
\[
B(\La_0)\ot \Bi^{\ot 2n-1}\ot \Bs
\simeq
B(\La_1)\ot\Bs,
\]
where $\blo$ is the highest weight element of $\Blo$ and
the subscript of $\fbj$ specifies to which crystal
the element belongs, $\fbj_s\in\Bs$.
The well definedness of $\psi$ follows from

\begin{lemma}
The definition of $\psi$ does not depend on the choice
of $n$ which satisfies the condition \ref{n}.
\end{lemma}
Proof\par
\noindent
It is sufficient to prove
\begin{eqnarray}
&&
\fbp\ot\fbm\ot
\psi_n(\fbj_s\ot b^\prime)
=
\psi_{n+2}(\fbj_s\ot\fbm\ot\fbp\ot b^\prime),
\nonumber
\end{eqnarray}
for $0\leq j\leq s-1$, $n\in {\bf Z}_{\geq 1}$ and any
$b^\prime\in B_1^{\ot n}$.
These equations follow from Lemma \ref{map}.
$\Box$

\begin{lemma}
The map $\psi$ commutes with the actions of
$\et_1$ and $\ft_1$.
\end{lemma}
Proof\par
\noindent
Let $B$ be the connected component, as a $\{1\}$-crystal,
of $\Bsm\ot\Bi$ which contains
$\fbosm\ot\fbp$. Then
\[
B=\{\fbjsm\ot\fbp| \hbox{ $0\leq j\leq s-1$}\}
\sqcup
\{\fbsmsm\ot\fbm\}
\]
and $B$ is isomorphic to $\Bs$ as a $\{1\}$-crystal by the map
\begin{eqnarray}
B&\ra&\Bs
\nonumber
\\
\fbjsm\ot\fbp&\mapsto&\fbjs
\hbox{ for $0\leq j\leq s-1$}
\nonumber
\\
\fbsmsm\ot\fbm&\mapsto&\fbss.
\nonumber
\end{eqnarray}

Let $\fbj_s\ot b\in \Bsm\ot \Blo$ and $n$ as above.
Now we shall describe $\psi$ as a composition of
several crystal morphisms from the connected component
of $\fbj_s\ot b$ as a $\{1\}$-crystal.
First of all
\begin{eqnarray}
\Bs\ot B(\La_0)
&\simeq&
\Bsm\ot B(\La_0)\ot\Bi^{\ot 2n}
\nonumber
\\
\fbjsm\ot b
&\mapsto&
\fbjsm\ot\bla\ot\fbp\ot b_{2n-1}\ot \cdots b_1=:\tilde{b}
\nonumber
\end{eqnarray}
is an isomorphism of classical crystals.
The crystal $B\ot\Bi^{\ot 2n-1}$ is a sub $\{1\}$-crystal
of $\Bsm\ot B(\La_0)\ot\Bi^{\ot 2n}$, by the map
\begin{eqnarray}
&&
\fbjsm\ot\fbox{$\epsilon$}\ra
\fbjsm\ot\blo\ot\fbox{$\epsilon$}\ot b_{2n-1}\ot\cdots\ot b_1.
\nonumber
\end{eqnarray}
The element $\tilde{b}$ is in this subcrystal.
Next
\[
B\ot\Bi^{\ot 2n-1}
\simeq
\Bs\ot\Bi^{\ot 2n-1}
\]
as a $\{1\}$-crystal as we already discussed.
We have the isomorphism of $\{0,1\}$-crystal
\[
\psi_{2n-1}:\Bs\ot\Bi^{\ot 2n-1}
\simeq
\Bi^{\ot 2n-1}\ot\Bs.
\]
Finally we have the injective $\{1\}$-crystal morphism
\begin{eqnarray}
\Bi^{\ot 2n-1}\ot\Bs
&\ra&
B(\La_0)\ot\Bi^{\ot 2n-1}\ot \Bs
\nonumber
\\
b'
&\mapsto&
\bla\ot b'.
\nonumber
\end{eqnarray}
It is easy to check that the map $\psi$ is the composition of the above maps.
 Since we can take sufficiently large $n$ such
that the condition \ref{n} holds for For $\fbjmsm\ot b$,
$\ft_1(\fbjmsm\ot b)$ and
$\et_1(\fbjmsm\ot b)$, $\psi$ is a $\{1\}$-crystal morphism.
$\Box$

\begin{lemma}
The map $\psi$ commutes with the action of
$\et_0$ and $\ft_0$.
\end{lemma}
Proof\par
\noindent
Let us define a map $\psit$ in the following manner.
For $\fbjsm\ot b\in \Bsm\ot B(\La_0)$, take $n\in{\bf Z}_{\geq0}$
such that
\[
b=(b_k)_{k=1}^{\infty},\quad
b_k=(-1)^k\hbox{ for $k\geq 2n+1$}.
\]
Then
\[
\psit(\fbjsm\ot b)=\blai\ot\psi_{2n}(\fbjps\ot b_{2n}\ot\cdots\ot b_1)
\]
through the isomorphism
\[
B(\La_1)\ot\Bi^{2n}\ot\Bs
\simeq
B(\La_1)\ot\Bs.
\]
In a similar manner to the $\psi$ case, we can check that
the definition of $\psit$ is independent of the choice of $n$.

\newtheorem{sublemma}{Sublemma}

\begin{sublemma}
$\psi=\psit$.
\end{sublemma}
Proof\par
\noindent
We use the above notations.
Take $n$ in such a way that satisfies the condition \ref{n}.
Then
\[
\psi(\fbjsm\ot b)=\psit(\fbjsm\ot b)
\]
is equivalent to
\[
\fbm\ot\psi_{2n-1}(\fbjs\ot b_{2n-1}\ot\cdots\ot b_1)
=
\psi_{2n}(\fbjps\ot\fbp\ot b_{2n-1}\ot\cdots\ot b_1)
\]
for $0\leq j\leq s-1$.
This follows from Lemma \ref{map}.
$\Box$
\par
Now the commutativity of $\psit$ and the action of
$\ft_0$ and $\et_0$ is similarly proved as before.
Namely let us set
\[
B'=\{\fbjsm\ot\fbm|\hbox{ $0\leq j\leq s-1$}\}
\sqcup
\{\fbo\ot\fbp\}.
\]
Then this constitutes, as a $\{0\}$-crystal,
a connected component of $\Bsm\ot\Bi$ isomorphic to
$\Bs$.
The map is given by
\begin{eqnarray}
B'&\ra&\Bs
\nonumber
\\
\fbjsm\ot\fbm&\mapsto&\fbjps
\hbox{ for $o\leq j\leq s-1$}
\nonumber
\\
\fbos\ot\fbp&\mapsto&\fbos.
\nonumber
\end{eqnarray}
Using this description
it is easy to show that the $\psit$ is described as a composition of
$\{0\}$ crystal morphisms from
any $\{0\}$-crystal connected component as before.
Hence the lemma is proved.
$\Box$

\begin{lemma}
$\psi$ is a bijection.
\end{lemma}
Proof\par
\noindent
We shall prove the injectivity first.
Suppose that
\[
\psi(\fbjsm\ot b)
=
\psi(\fbjprsm\ot b').
\]
By the definition of $\psi$ this is equivalent to
\[
\psi_{2n-1}(\fbjs\ot b_{2n-1}\ot\cdots\ot b_1)
=
\psi_{2n-1}(\fbjprs\ot b_{2n-1}^\prime\ot\cdots\ot b_1^\prime),
\]
for sufficiently large $n$.
Since $\psi_{2n-1}$ is bijective, we have
\[
j=j',\quad
b_k=b_k'\hbox{ for $1\leq k\leq 2n-1$}
\]
which means $b=b^\prime$.
The surjectivity easily follows from Lemma \ref{map}:
\begin{eqnarray}
&&
\psi_1^{-1}(\fbp\ot\fbjps)
=
\fbjs\ot\fbm
\hbox{ for $0\leq j\leq s-1$}
\nonumber
\\
&&
\psi_1^{-1}(\fbm\ot\fbjs)
=\fbjps\ot \fbp\hbox{ for $0\leq j\leq s-1$},
\nonumber
\\
&&
\psi_1^{-1}(\fbm\ot\fbss)
=\fbss\ot \fbm.
\nonumber
\end{eqnarray}
$\Box$\par
\noindent
This lemma complets the proof of theorem \ref{mainth}.

\section{Existence of new type of vertex operators}\label{EVO}
In this section we shall prove the existence of new types of
$q$-vertex operators, one type of which is conjectured to induce the crystal
isomorphisms in section \ref{crystal}.
For non-zero complex numbers $z_1,\cdots,z_k$ and $(i,j)\in \{0,1\}^2$
let us define the $F[z_1^{\pm1},\cdots,z_k^{\pm1}]$ module
by
\begin{eqnarray}
&&
H^{n_1\cdots n_k}_{z_1\cdots z_k}(i,j)
=\{v\in (V_{n_1})_{z_1}\ot\cdots\ot(V_{n_k})_{z_k}|
wt(v)=\La_i-\La_j,
e_l^{<h_l,\La_j>+1}v=0\hbox{ for $l=0,1$}\}.
\nonumber
\end{eqnarray}
Our aim here is to prove

\begin{theor}\label{vo}
\begin{description}
\item[(1)]
$H^{n,m}_{z_2,q^{-3}z_1}(i,i+1)$,
$H^{n,n}_{q^2z_1,z_2}(i,i)$
and
$H^{n+1,1,n}_{z_1,z_2,q^{-3}z_1}(i,i+1)$
are free $F[z_1^{\pm1},z_2^{\pm1}]$ modules and their ranks
are given by
\begin{eqnarray}
&&
\hbox{rank} H^{n,m}_{z_2,q^{-3}z_1}(i,i+1)=\delta_{|n-m|,1}\delta_{z_1,z_2},
\nonumber
\\
&&
\hbox{rank} H^{n,n}_{q^2z_1,z_2}(i,i)=\delta_{z_1,z_2},
\nonumber
\\
&&
\hbox{rank} H^{n+1,1,n}_{z_1,z_2,q^{-3}z_1}(i,i+1)=1.
\nonumber
\end{eqnarray}
\item[(2)]
There are isomorphisms of $F[z_1^{\pm1},z_2^{\pm1}]$ modules
\begin{eqnarray}
&&
\hbox{Hom}_{F^\prime\ot U'}
\big(
(\Vm)_{z_1}\ot V(\La_i), V(\La_{i+1})\hat{\ot}(\Vn)_{z_2}
\big)
\simeq H^{n,m}_{z_2,q^{-3}z_1}(i,i+1),
\nonumber
\\
&&
\hbox{Hom}_{F^\prime\ot U'}
\big(
(\Vn)_{z_1}\ot V(\La_i),(\Vn)_{z_2}\hat{\ot}V(\La_i)
\big)
\simeq H^{n,n}_{q^2z_1,z_2}(i,i),
\nonumber
\\
&&
\hbox{Hom}_{F^\prime\ot U'}
\big(
(\Vn)_{z_1}\ot V(\La_i), V(\La_{i+1})\hat{\ot}(\Vnp)_{z_1}
\ot(V_1)_{z_2}
\big)
\simeq H^{n+1,1,n}_{z_1,z_2,q^{-3}z_1}(i,i+1),
\nonumber
\end{eqnarray}
where $F^\prime=F[z_1^{\pm1}]$.
\end{description}
\end{theor}

\begin{cor}
\begin{eqnarray}
&&
\hbox{Hom}_{U'}
\big(
V(\La_i), (\Vn)_{q^2z}\hat{\ot}V(\La_{i+1})\hat{\ot}(\Vnp)_z
\big)
\nonumber
\\
&&
\simeq
\hbox{Hom}_{F[z^{\pm1}]\ot U'}
\big(
V(\La_i)\ot(\Vnp)_z, (\Vn)_{z}\hat{\ot}V(\La_{i+1})
\big)
\simeq
F[z^{\pm1}].
\nonumber
\end{eqnarray}
\end{cor}

Let us prove (1). Other cases are similarly proved.
Note that
\begin{eqnarray}
&&\hbox{Hom}_{U'}
\big(
(\Vm)_{z_1}\ot V(\La_i), V(\La_{i+1}) \hat{\ot} (\Vn)_{z_2}
\big)
\nonumber
\\
&&\simeq
\hbox{Hom}_{U'}
\big(
V(\La_i), (\Vm)_{q^2z_1}\hat{\ot} V(\La_{i+1}) \hat{\ot} (\Vn)_{z_2}
\big).
\nonumber
\end{eqnarray}
Let $\Ub$ be the subalgebra of $U'$ generated by
$e_i$, $t_i$ $(i=0,1)$. Then we have
\begin{eqnarray}
&&
\hbox{Hom}_{U'}
\big(
V(\La_i), (\Vm)_{q^2z_1}\hat{\ot} V(\La_{i+1}) \hat{\ot} (\Vn)_{z_2}
\big)
\nonumber
\\
&&
\simeq
\hbox{Hom}_{\Ub}
\big(
\qquli, (\Vm)_{q^2z_1}\hat{\ot} V(\La_{i+1}) \hat{\ot} (\Vn)_{z_2}
\big)
\nonumber
\\
&&
\simeq
\hbox{Hom}_{\Ub}
\big(
 V(\La_{i+1})^{\ast a}\ot(\Vm)_{z_1}\ot\qquli,
  (\Vn)_{z_2}
\big).
\nonumber
\end{eqnarray}
Here we used the following lemma which can be proved
in a similar way to that in \cite{DJO}.

\begin{lemma}
Take any $i$ and fix it. Let $u\in(V_n)_{z}\hot \Vli
\hot(V_m)_{z}$ be a weight vector of $t_i$.
If $u$ satisfies $e_i^lu=0$ for some $l$, then $f_i^Nu=0$ for some $N$.
\end{lemma}

\begin{lemma}
There is an isomorphism of $\Ub$-modules,
\[
(\Vn)_{z}\ot Fu_{\La_i}
\simeq
Fu_{\La_i}\ot(\Vn)_{q^{-1}z}
\]
given by the map
\[
\vnj\ot\uli
\ra
q^{-ji}\uli\ot\vnj.
\]
\end{lemma}
This lemma implies
\begin{eqnarray}
&&
\hbox{Hom}_{\Ub}
\big(
 V(\La_{i+1})^{\ast a}\ot(\Vm)_{z_1}\ot\qquli,
  (\Vn)_{z_2}
\big)
\nonumber
\\
&&
\simeq
\hbox{Hom}_{\Ub}
\big(
 V(\La_{i+1})^{\ast a}\ot\qquli,
  (\Vn)_{z_2}\ot(\Vm)_{q^{-3}z_1}
\big)
\nonumber
\\
&&
\simeq H^{n,m}_{z_2,q^{-3}z_1}(i,i+1).
\label{hom}
\end{eqnarray}
Let us write explicitly the conditions satisfied by the vector $v$
of $H^{n,m}_{z_2,q^{-3}z_1}(i,i+1)$ according to $i=0,1$;
\begin{eqnarray}
&&
wt(v)=\La_0-\La_1,
\quad
e_1^2v=e_0v=0,
\hbox{ if $i=0$}
\label{io}
\\
&&
wt(v)=\La_1-\La_0,
\quad
e_1v=e_0^2v=0,
\hbox{ if $i=1$}.
\label{ii}
\end{eqnarray}

Let us determine the vectors which satisfies the condition
\ref{io} and \ref{ii}.
Note first that the condition  \ref{io} or \ref{ii} implies
$|n-m|=1$. In fact the vector satisfying \ref{io} or \ref{ii}
must lie
in the two dimensional irreducible representation of $\Upi$.

Let $\wj$ be the highest weight vectors of
$\Vn\ot\Vm$ with the weight $(n+m-2j)\La_1$
as a $\uqi$-module.
They are explicitly given by
\begin{eqnarray}
&&
\wj=\sum_{k=0}^{j}c^{(j)}_k(n) v^{(n)}_k\ot v^{(m)}_{j-k},
\label{hwtii}
\\
&&
c^{(j)}_k(n)=(-1)^kq^{k(n+1-k)}\bnjk,
\quad c^{(0)}_0(n)=1.
\label{hwti}
\end{eqnarray}
\par
\noindent
(1.1) $i=0$ and $n=m+1$ case
\par
\noindent
The vector satisfying the condition \ref{io} is proportional to
$f_1w_{m}$.
Let us calculate $e_0f_1w_{m}$ in the tensor product
$(\Vmp)_{z_2}\ot(\Vm)_{q^{-3}z_1}$. The result is
\begin{eqnarray}
&&
f_1w_{m}
=\sum_{k=1}^{m+1}
c_{k-1}^{(m)}(m+1)q^{m-2k+3}
(q^{-1}[m+2-k]-[m+1-k])v^{(m+1)}_k\ot v^{(m)}_{m+1-k}
\nonumber
\\
&&
e_0f_1w_{m}
=(z_2-z_1)
\sum_{k=2}^{m+1}
c_{k-2}^{(m)}(m+1)q^{-k+2}[m+2-k]v^{(m+1)}_k\ot v^{(m)}_{m-k+2}.
\nonumber
\end{eqnarray}
Hence $e_0v=0$ is equivalent to $z_1=z_2$.

\par
\noindent
(1.2) $i=0$ and $m=n+1$ case
\par
\noindent
The vector satisfying the condition \ref{io} is proportional to
$f_1w_{n}$. We have
\begin{eqnarray}
&&
f_1w_n=
\sum_{k=0}^nc^{(n)}_k(n)(-q^{-1}[k]+[k+1])\vnk\ot v^{(n+1)}_{n+1-k},
\nonumber
\\
&&
e_0f_1w_n=(z_1-z_2)
\sum_{k=1}^n c^{(n)}_{k}(n) q^{-n+3(k-1)}[k]
\vnk\ot v^{(n+1)}_{n+2-k}.
\nonumber
\end{eqnarray}
Hence $e_0f_1w_n=0$ is equivalent to $z_1=z_2$.

\par
\noindent
(1.3) $i=1$ and $n=m+1$ case
\par
\noindent
The vector satisfying the condition \ref{ii} is proportional
to $w_m$. Then
\begin{eqnarray}
&&
e_0^2w_m=
(z_1-z_2)
\sum_{k=2}^{m+1}
c^{(m)}_{k-2}(m+1)
(z_1[m+1-k]-z_2[m+3-k])
v^{(m+1)}_k\ot v^{(m)}_{m+2-k}.
\nonumber
\end{eqnarray}
Therefore $e_0^2w_m=0$ if and only if $z_1=z_2$.

\par
\noindent
(1.4) $i=1$ and $m=n+1$ case
\par
\noindent
The vector satisfying the condition \ref{ii} is proportional
to $w_n$. Then we have
\begin{eqnarray}
&&
e_0^2w_n=
q^{-6}(z_1-z_2)
\sum_{k=1}^{n}
c^{(n)}_{k}(n) q^{-2n+4k}
(z_1[k+1]-z_2[k-1])
v^{(n)}_k\ot v^{(n+1)}_{n+2-k}.
\nonumber
\end{eqnarray}
Consequently $e_0^2w_n=0$ iff $z_1=z_2$.
$\Box$

By theorem \ref{vo} there are uniquely determined $\Up$
intertwinwers
\begin{eqnarray}
&&
\phinnp(z):
(\Vn)_z\ot V(\La_i)\ra \Vlip\hot (\Vnp)_z,
\nonumber
\\
&&
\phiunnp(z):
\Vli\ra (\Vn)_{q^2z}\ot \Vlip\hot (\Vnp)_z,
\nonumber
\\
&&
\udphinnp(z):
\Vli\ot (\Vnp)_z\ra (\Vn)_z\hot \Vlip,
\nonumber
\end{eqnarray}
under the normalizations
\begin{eqnarray}
&&
<\udlip,\phinnp(z)(\vnj\ot\uli)>
=v^{(n+1)}_{1-i+j},
\nonumber
\\
&&
<v^{(n)\ast}_j\ot \udlip,\phiunnp(z)(\uli)>
=(-1)^jq^{j(n+1-j)}\bnnj v^{(n+1)}_{n+1-j-i},
\nonumber
\\
&&
<\udlip,\udphinnp(z)(\uli\ot\vnpj)>
=q^{(2i-1)j-i}
{
[j]^i[n+1-j]^{1-i}
\over
[n+1]^{1-i}
}
v^{(n)}_{j-i}
\quad
(i\leq j\leq n+i).
\nonumber
\end{eqnarray}
In the following we also call those intertwiners simply
vertex operators.

\newtheorem{rem}{Remark}
\begin{rem}
I conjecture that the vertex opertator $\phinnp(z)$
preserves the crystal lattice and induces the
isomorphism of crystals in section \ref{crystal}.
Some part of Miki's conjecture\cite{M} is a special case of
this conjecture.
\end{rem}

\section{Fusion of representations}
Let us briefly
recall the fusion construction
of representations and $R$-matrices in order to fix notations.
Let $M_i$ be the trivial representation in the tensor product
$(V_1)_{q^{-2(i-1)}z}\ot(V_1)_{q^{-2i}z}$ for $1\leq i\leq n$.
Explicitly $M_i$ is written
\[
M_i=F\cdot(v^{(1)}_0\ot v^{(1)}_1-qv^{(1)}_1\ot v^{(1)}_0).
\]
Let us set $N_i=(V_1)_{z}\ot\cdots\ot M_i\ot\cdots\ot(V_1)_{q^{-2n}z}$
and
\begin{eqnarray}
&&
W_{n+1}(z)=
(V_1)_{z}\ot\cdots\ot(V_1)_{q^{-2n}z}/\sum_{i=1}^nN_i,
\nonumber
\\
&&\tilde{W}_{n+1}(z)=\Up(v^{(1)}_0\ot\cdots\ot v^{(1)}_0)
\hookrightarrow (V_1)_{q^{-2n}z}\ot\cdots\ot (V_1)_{z}
\nonumber
\end{eqnarray}
Then the following proposition is well known.

\begin{prop}\label{repfusion}
$W_{n+1}(z)\simeq \tilde{W}_{n+1}(z)\simeq
(V_{n+1})_{q^{-n}z}.$
\end{prop}

In order to describe the isomorphism explicitly we
shall introduce the following definitions.

\begin{defi}
\begin{description}
\item[(1)]
$(\ep_1,\cdots,\ep_n)$ is of type $j$ if and only if
$\sharp\{k|\ep_k=1\}=j$.
\item[(2)]
For $(\ep_1,\cdots,\ep_n)$ let us define its inversion number
by
\begin{eqnarray}
&&
\inv(\ep_1,\cdots,\ep_n)=
\sum_{i:\ep_i=1}\sharp\{k|\ep_k=0,k<i\}.
\nonumber
\end{eqnarray}
\end{description}
\end{defi}

Then the isomorphism is given by
\begin{eqnarray}
W_{n+1}(z)&\ra&(V_{n+1})_{q^{-n}z}
\nonumber
\\
v^{(1)}_{\ep_1}\ot\cdots\ot v^{(1)}_{\ep_n}
&\mapsto&
q^{\inv(\ep_1,\cdots,\ep_n)}v^{(n+1)}_j
\nonumber
\end{eqnarray}
for $(\ep_1,\cdots,\ep_n)$ of type $j$.

Let us give the description of $\tilde{W}_{n+1}$
in terms of R-matrix for the later use.
Let $\brc({z_1\over z_2})$ be the $\Up$ linear morphism
$(V_1)_{z_1}\ot(V_1)_{z_2}\ra(V_1)_{z_2}\ot(V_1)_{z_1}$
such that $\brc({z_1\over z_2})(v^{(1)\ot 2}_0)=v^{(1)\ot 2}_0$.
Consider the composition
$\brc_{n+1}(z)=
\brc({z_n\over z_{n+1}})\cdots
\brc({z_1\over z_{n+1}})\cdots
\brc({z_1\over z_2})$
at $z_j=q^{-2(j-1)}z (1\leq j\leq n+1)$.
Then it is well known (and easily proved) that

\begin{prop}\label{ImR}
\begin{eqnarray}
&&
Im\brc_{n+1}(z)=\tilde{W}_{n+1}(z),
\quad
Ker\brc_{n+1}(z)=\sum_{k=1}^nN_k,
\nonumber
\end{eqnarray}
\end{prop}

\subsection{Fusion of the R-matrix}
Let $\brc_{n+1,1}({z\over w})=
\brc({q^nz\over w})\brc({q^{n-2}z\over w})
\cdots\brc({q^{-n}z\over w})$ be the $\Up$ intertwiner
$(V_1)_{q^nz}\ot\cdots(V_1)_{q^{-n}z}\ot(V_1)_{w}
\ra(V_1)_{w}\ot(V_1)_{q^nz}\ot\cdots(V_1)_{q^{-n}z}$.
Then

\begin{prop}
$\brc_{n+1,1}({z\over w})$ induces the $\Up$
linear map $W_{n+1}(q^nz)\ot(V_1)_{w}\ra (V_1)_{w}\ot W_{n+1}(q^nz)$
such that the following diagram is commutative:
\[
\begin{array}{ccc}
(V_1)_{q^nz}\ot\cdots(V_1)_{q^{-n}z}\ot(V_1)_{w}
&
\stackrel{\brc_{n+1,1}({z\over w})}{\ra}
&
(V_1)_{w}\ot(V_1)_{q^nz}\ot\cdots(V_1)_{q^{-n}z}
\\
\downarrow
&
{}
&
\downarrow
\\
W_{n+1}(q^nz)\ot(V_1)_{w}
&
\ra
&
(V_1)_{w}\ot W_{n+1}(q^nz).
\end{array}
\]
Here the downarrows are the natural projections.
\end{prop}
\par
\noindent
Proof:
It is sufficient to prove
\begin{eqnarray}
&&
\brc_{n+1,1}({z\over w})(N_j\ot (V_1)_{w})\subset
(V_1)_{w}\ot\sum_{k=1}^nN_k.
\nonumber
\end{eqnarray}
By Proposition \ref{ImR} this is equivalent to
\begin{eqnarray}
&&
\brc_{n+1}(q^nz)\brc_{n+1,1}({z\over w})(N_j\ot (V_1)_{w})=0
\nonumber
\end{eqnarray}
which follows from the Yang-Baxter equation.
$\Box$

We use the same symbol for $\brc_{n+1,1}({z\over w})$
for the induced map.
Then, explicitly, $\brc_{n+1,1}(z)$ is given by
\[
\brc_{n,1}(z)
\left[
\begin{array}{c}
v^{(n)}_k\ot v^{(1)}_1
\\
v^{(n)}_{k+1}\ot v^{(1)}_0
\end{array}
\right]
=
{1\over 1-q^{n+1}z}
\left[
\begin{array}{cc}
-q^{k+1}z+q^{n-k}
&
(1-q^{2n-2k})z
\\
1-q^{2k+2}
&
-q^{n-k}z+q^{k+1}
\end{array}
\right]
\left[
\begin{array}{c}
v^{(1)}_1\ot v^{(n)}_k
\\
v^{(1)}_0\ot v^{(n)}_{k+1}
\end{array}
\right].
\nonumber
\]

\section{Fusion of $q$-vertex operators}\label{FVO}
In this section we shall give a construction
of the vertex operator $\phinnp(z)$ in terms of
$\Phi(z)$ and $\Psi(z)$.
The idea is to consider the composition
\[
\begin{array}{llc}
\Vli
&
\ra
&
(V_1)_{q^{n+1}z}\ot\cdots\ot(V_1)_{q^{-n+3}z}\ot\Vlip\ot
\ot (V_1)_{q^nz}\ot\cdots\ot(V_1)_{q^{-n}z}
\\
{}&{}&\downarrow
\\
{}&{}&
(V_n)_{q^{-n+2}z}\ot\Vlip\ot(V_{n+1})_{q^{-n}z}.
\end{array}
\]
For the sake of simplicity hereafter we omit writting
the the symbol $\hat{}$ of the extended tensor product.
The vertical arrow is the $\Up$-linear projection
defined in Proposition \ref{repfusion}.
Unfortunately the composition of vertex operators $\Phi$ and $\Psi$
which gives the horizontal arrow is not well defined in general.
So we must carefully proceed in the following manner.
Let us define the operator $\ozu$,
$({\bf z},{\bf u})\in {\bf C}^{\ast n+1}\times{\bf C}^{\ast n}$,
acting on $\Vli$ by
\begin{eqnarray}
&&
O(z_1,\cdots,z_{n+1}|u_n,\cdots,u_{1})={1\over f}
\Phi(z_1)\cdots\Phi(z_{n+1})
\Psi(u_n)\cdots\Psi(u_1),
\nonumber
\end{eqnarray}
where ${\bf C}^\ast=\{z\in {\bf C}|z\neq0\}$ and
\begin{eqnarray}
&&
f(z_1,\cdots,z_{n+1}|u_n,\cdots,u_{1})=
\prod_{j<k}{({q^2z_k\over z_j})_\infty\over({q^4z_k\over z_j})_\infty}
\prod_{j>k}{({u_k\over u_j})_\infty\over({q^2u_k\over u_j})_\infty}
\prod_{j,k}{({qu_j\over z_k})_\infty\over({u_j\over qz_k})_\infty}.
\nonumber
\end{eqnarray}
The operator $\ozu$ satisfies, on $\Vli$, the symmetry relations
\begin{eqnarray}
&&
\big({z_j\over z_{j+1}}\big)^{-1+\overline{i+j}}
\brc({z_j\over z_{j+1}})\ozu=O(\sigma_j{\bf z}|{\bf u}),
\nonumber
\\
&&
\big({u_j\over u_{j+1}}\big)^{\overline{i+j}}
\brc({u_j\over u_{j+1}})\ozu=O({\bf z}|\sigma_j{\bf u}),
\nonumber
\end{eqnarray}
where $\sigma_j$ is the permutation exchanging
$z_j,z_{j+1}$ or $u_j,u_{j+1}$ and $\overline{k}=0,1$ according to
$k$ is even or odd.
Let
\begin{eqnarray}
&&
Pr(z)_{jk}:(V_1)_{z_j}\ot(V_1)_{z_k}\ra V_2,
\nonumber
\\
&&
Pr(u)_{jk}:(V_1)_{u_j}\ot(V_1)_{u_k}\ra V_2,
\nonumber
\\
&&
Pr(z):(V_1)_{z_1}\ot\cdots\ot(V_1)_{z_{n+1}}\ra V_{n+1},
\nonumber
\\
&&
Pr(u):(V_1)_{u_1}\ot\cdots\ot(V_1)_{u_{n}}\ra V_{n},
\nonumber
\end{eqnarray}
be the $\uqi$-linear projection normalized as
\begin{eqnarray}
&&
Pr(z)_{jk}(v^{(1)\ot2}_0)=v^{(2)}_0,
\nonumber
\\
&&
Pr(z)(v^{(1)\ot n+1}_0)=v^{(n+1)}_0
\nonumber
\end{eqnarray}
and similarly for $Pr(u)_{jk}$, $Pr(u)$.
Since $Pr(z)$ and $Pr(u)$ is determined uniquely under these
normalizations, we have, for $j<k$
\begin{eqnarray}
&&
Pr(z)=Pr(z)\brc({z_j\over z_{k-1}})\cdots\brc({z_j\over z_{j+1}}).
\label{prcom}
\end{eqnarray}
The $Pr(z)$ in the right hand side is the $\uqi$ linear
projection
\begin{eqnarray}
&&
(V_1)_{z_1}\ot\cdots\ot(V_1)_{z_j}\ot(V_1)_{z_k}\ot
\cdots\ot(V_1)_{z_{n+1}}\ra V_{n+1}.
\nonumber
\end{eqnarray}
To simplify the notations we use the same symbol $Pr(z)$
althought the space acted by it is different from that
of $Pr(z)$ in the left hand side.
Note that there is an $\uqi$-linear projection such that
\[
\begin{array}{ccc}
(V_1)_{z_1}\ot\cdots\ot(V_1)_{z_j}\ot(V_1)_{z_{j+1}}\ot
\cdots\ot(V_1)_{z_{n+1}}
&
\stackrel{Pr(z)_{jk}}{\ra}
&
(V_1)_{z_1}\ot\cdots\ot V_2\ot\cdots\ot(V_1)_{z_{n+1}}
\\
\downarrow
&
{}
&
\downarrow Pr(z)^{jk}
\\
V_{n+1}
&
=
&
V_{n+1}
\end{array}
\]
is a commutative diagram.

\begin{prop}\label{regular}
\begin{description}
\item[(1)]
The operator $\ozu$ has poles at most simple at
$z_j=q^2z_k$ $(j< k)$ and $u_j=q^2u_k$ $(j< k)$.
\item[(2)]
The operator $Pr(z)Pr(u)\ozu$ is regular at
$z_j=q^2z_k$ $(j< k)$ and $u_j=q^2u_k$ $(j< k)$.
\end{description}
\end{prop}
\noindent
Proof:(1) The integral formula of $<\La_{i+1}|\ozu|\La_i>$
(appendix?) implies that $\ozu$ has poles at most at
$z_j=q^2z_k$ $(j< k)$, $u_j=q^2u_k$ $(j< k)$ and
$u_j=qz_k,q^{3}z_k$ for any $j,k$.
Because there is a possibility to occur a pinch of the integration
pathes only in those cases. Moreover it is easy to prove that
these poles are at most simple.
Hence it is sufficient to prove that there are no poles
at $u_j=qz_k,q^{3}z_k$ for any $j,k$.
But again this follows easily from the integral formula
of $<\La_{i+1}|\ozu|\La_i>$ by the following reason.
Consider a component of $<\La_{i+1}|\ozu|\La_i>$.
Let us decompose each integral as
\begin{eqnarray}
&&
\int_{C_d}{d\xi_d\over 2\pi i}=
\int_{C_0}{d\xi_d\over 2\pi i}
+\sum_{j=1}^dRes_{\xi_d=u_j},
\quad
\int_{C_a}{dw_a\over 2\pi i}=
\int_{C_\infty}{dw_a\over 2\pi i}
-\sum_{j=1}^aRes_{w_a=q^2z_j},
\nonumber
\end{eqnarray}
where $C_0$, $C_\infty$ are the small circles around
$0$, $\infty$ going anti-clockwise and clockwise
direction respectively.
Here, for the sake of simplicity, we omit writing the integrands.
Then the integral which we consider now
is a sum of terms of the form
\begin{eqnarray}
&&
\prod_{d\in D_1}\int_{C_0}{d\xi_d\over 2\pi i}
\prod_{d\in A_1}\int_{C_\infty}{dw_a\over 2\pi i}
Res_{w_{a_r}=q^2z_{j_r}}\cdots Res_{w_{a_1}=q^2z_{j_1}}
Res_{\xi_{d_l}=u_{i_l}}\cdots Res_{\xi_{d_1}=u_{i_1}},
\nonumber
\end{eqnarray}
where $D_1$ and $A_1$ is a subset of $\{a\}$ and $\{d\}$ respectively.
Since there is a term
$\prod_{a<a'}(1-{w_{a'}\over w_a})\prod_{d<d'}(1-{\xi_{d'}\over \xi_d})$
in the numerator of the integrand, we can assume that
$j_{p_1}\neq j_{p_2}(p_1\neq p_2)$,
$i_{l_1}\neq i_{l_2}(l_1\neq l_2)$.
In $Res_{\xi_{d_l}=u_{i_l}}\cdots Res_{\xi_{d_1}=u_{i_1}}$
the possible poles at $w_a=qu_{i_k}$ are cancelled out with
$\prod_a\prod_{l=1}^{n}(1-{qu_l\over w_a})$.
Hence after taking residues in $w_{a_p}s$, there does not appear
poles at $u_j=qz_k$.
Since there is the term $\prod_d\prod_{j=1}^{n+1}(1-{\xi_d\over q^3z_j})$
in the numerator,
the poles at $u_{i_p}=q^3z_{j_k}$ which appear after taking
$Res_{w_{a_r}=q^2z_{j_r}}\cdots Res_{w_{a_1}=q^2z_{j_1}}$
are also cancelled out.
Finally in the remaining integral
$\prod_{d\in D_1}\int_{C_0}{d\xi_d\over 2\pi i}
\prod_{d\in A_1}\int_{C_\infty}{dw_a\over 2\pi i}$
there does not occur pinches of the integral contours
at $u_j=qz_k,q^3z_k$. Hence it has no singularities there.
\par
\noindent
(2): Let us consider the composition
\begin{eqnarray}
&&
\Phi(z_1)\Phi(z_2):
\Vli\ra\Vhlip\ot (V_1)_{z_1}\ot(V_1)_{z_2}.
\nonumber
\end{eqnarray}
By the explicit formula of $<\La_{i+1}|\ozu|\La_i>$ (\cite{DFJMN})
$\Phi(z_1)\Phi(z_2)$ is regular at $z_1=q^2z_2$.
Since there is no non-zero $\Up$ intertwiner
$
\Vli\ra\Vhlip\ot (V_2)_z,
$
we have
$
Pr(z)_{12}\Phi(q^2z_2)\Phi(z_2)=0.
$
Hence
\begin{eqnarray}
&&
Res_{z_j=q^2z_{j+1}}{1\over f}Pr(z)_{jj+1}\Phi(z_j)\Phi(z_{j+1})=0
\label{twocproj}
\end{eqnarray}
for any $1\leq j\leq n$.
Using the commutation relations of the vertex operators
$\Phi(z)$ and the relations \ref{prcom}, \ref{twocproj}
\begin{eqnarray}
&&
Res_{z_j=q^2z_k}Pr(z)\ozu
\nonumber
\\
&&
=
Res_{z_j=q^2z_k}
\prod_{l=j}^{k-2}
({z_j\over z_l})^{-1+\overline{i+l}}
Pr(z)
\brc({z_j\over z_{j+1}})^{-1}
\cdots\brc({z_j\over z_{k-1}})^{-1}
O(z_1,\cdots,z_j,z_k,\cdots,z_{n+1}|{\bf u})
\nonumber\\
&&
=\prod_{l=j}^{k-2}
({q^2z_k\over z_l})^{-1+\overline{i+l}}
Res_{z_j=q^2z_k}
Pr(z)
O(z_1,\cdots,z_j,z_k,\cdots,z_{n+1}|{\bf u})
\nonumber
\\
&&
=\prod_{l=j}^{k-2}
({q^2z_k\over z_l})^{-1+\overline{i+l}}
Pr(z)^{jk}
Res_{z_j=q^2z_k}
Pr(z)_{jk}
O(z_1,\cdots,z_j,z_k,\cdots,z_{n+1}|{\bf u})
\nonumber
\\
&&
=0.
\nonumber
\end{eqnarray}
Hence $Pr(z)\ozu$ is regular at $z_j=q^2z_k$ $(j< k)$.
We can similarly prove that
$Pr(u)\ozu$ is regular at $u_j=q^2u_k$ $(j< k)$.
$\Box$

\begin{defi}[Fused vertex operator]
\begin{eqnarray}
\oz&=&Pr(z)Pr(u)\ozu
|_{z_j=q^{-2j+2}z (1\leq j\leq n+1),
u_k=q^{-2k+3}z (1\leq k\leq n)}
\nonumber
\\
&=&
\sum_{j,k}
\vnj\ot\oz_{jk}\ot\vnpk.
\nonumber
\end{eqnarray}
\end{defi}

\begin{theor}
\begin{description}
\item[(1)]
The operator $\oz$ is not zero as a linear map.
\item[(2)]
$\oz$ is a $\Up$-linear map
\begin{eqnarray}
&&
\Vli\ra(V_n)_{q^{-n+2}z}\ot\Vlip\ot(V_{n+1})_{q^{-n}z}.
\nonumber
\end{eqnarray}
\end{description}
\end{theor}
\par
\noindent
Proof\par
\noindent
(1): The integral formula of $<\La_{i+1}|\ozu|\La_i>$
implies (see \ref{inti}, \ref{intii})
\begin{eqnarray}
&&
<\La_1|\oz_{0,n+1}|\La_0>=
(-1)^{[{n\over2}](n-1)}
(-q)^{{n(n-2)\over4}-{3\over8}(1-(-1)^n)}
\nonumber
\\
&&
<\La_0|\oz_{n,0}|\La_1>=
(-1)^{[{n\over2}](n-1)}
(-q)^{{n\over12}(8n^2-15n+22)+{3\over8}(1-(-1)^n)}
z^{-{n(n-1)\over2}}.
\nonumber
\end{eqnarray}
Hence $\oz$ is not zero as a linear map.
\par
\noindent
(2): By definition $\oz$ is $\uqi$-linear.
Therefore it is sufficient to prove that $\oz$
commutes with the actions of $e_0$ and $f_0$.

Let us prove the commutativity of $\oz$ with $e_0$.
The case of $f_0$ is similarly proved.
The intertwining properties of $\ozu$ implies
\begin{eqnarray}
<v^\prime|\ozu|e_0v>=&&
(e_0\ot 1)<v^\prime|\ozu|v>
+(t_0\ot 1)<v^\prime e_0|\ozu|v>
\nonumber
\\
&&
+(t_0\ot e_0)<v^\prime t_0|\ozu|v>
\nonumber
\end{eqnarray}
for any $v\in \Vli$, $v^\prime\in\Vlip$.
It is sufficient to prove, modulo
$\sum N_j\ot (V_1)_{z_1}\ot\cdots(V_1)_{z_{n+1}}+
(V_1)_{u_1}\ot\cdots(V_1)_{u_{n+1}}\ot \sum N_j$, that
\begin{eqnarray}
&&
Pr(u)Pr(z)(e_0\ot 1)<v^\prime|\ozu|v>
=(e_0\ot 1)Pr(u)Pr(z)<v^\prime|\ozu|v>,
\label{eqi}
\\
&&
Pr(u)Pr(z)(t_0\ot 1)<v^\prime e_0|\ozu|v>
=(t_0\ot 1)Pr(u)Pr(z)<v^\prime e_0|\ozu|v>,
\label{eqii}
\\
&&
Pr(u)Pr(z)(t_0\ot e_0)<v^\prime t_0|\ozu|v>=
(t_0\ot e_0)Pr(u)Pr(z)<v^\prime t_0|\ozu|v>,
\label{eqiii}
\end{eqnarray}
at $z_j=q^{-2(j-1)}z (1\leq j\leq n+1)$,
$u_j=q^{-2(j-1)+1}z (1\leq j\leq n)$.
It is easily proved, using proposition \ref{regular} (2),
that the left hand sides of equations \ref{eqi}-\ref{eqiii}
are regular at $z_j=q^2z_k(j< k)$, $u_j=q^2u_k(j< k)$.
Hence we can specialize variables as above.

Since $t_0$ acts on $(V_1)_{u_1}\ot\cdots\ot(V_1)_{u_n}$ as $t_1^{-1}$
and $Pr(u)$ is $\uqi$ linear, \ref{eqii} holds.
Let us prove the equation \ref{eqi}.
According as the decompositions
$(V_1)_{u_1}\ot\cdots\ot(V_1)_{u_n}\simeq V_n\oplus\sum N_j$,
$(V_1)_{z_1}\ot\cdots\ot(V_1)_{z_{n+1}}\simeq V_{n+1}\oplus\sum N_j$ as
$\uqi$ modules, we write
\begin{eqnarray}
&&
<v^\prime|\ozu|v>=(A+A^\prime)\ot(B+B^\prime),
\nonumber
\\
&&
A\in V_n,
\quad
A^\prime\in \sum N_j,
\quad
B\in V_{n+1},
\quad
B^\prime\in \sum N_j.
\nonumber
\end{eqnarray}
Then
\begin{eqnarray}
&&
Pr(u)Pr(z)(e_0\ot 1)<v^\prime|\ozu|v>-
(e_0\ot 1)Pr(u)Pr(z)<v^\prime|\ozu|v>
\nonumber
\\
&&
=
(Pr(u)e_0A-e_0A)\ot B+Pr(u)e_0A^\prime\ot B.
\nonumber
\end{eqnarray}
Since $Pr(u)e_0A-e_0A\equiv0\hbox{ mod $\sum N_j$}$,
it is sufficient to prove
\begin{eqnarray}
&&
Pr(u)e_0A^\prime\ot B=0
\hbox{ at \condzu}.
\label{preqi}
\end{eqnarray}

\begin{lemma}
$Pr(u)e_0A^\prime\ot B$ is regular at $z_j=q^2z_k(j< k)$,
$u_j=q^2u_k(j< k)$.
\end{lemma}
\par
\noindent
Proof: Since $Pr(u)(e_0\ot 1)<v^\prime|\ozu|v>$ is regular at
$u_j=q^2u_k(j< k)$ as we already mentioned,
$Pr(u)(e_0\ot 1)(A+A^\prime)\ot B$
is regular at the same place.
On the other hand $(A+A^\prime)\ot B$ is regular at
$z_j=q^2z_k(j< k)$ and $A\ot B$ is regular at
$z_j=q^2z_k(j< k)$, $u_j=q^2u_k(j< k)$,
by proposition \ref{regular} (2).
Hence $Pr(u)e_0A^\prime\ot B$ is regular at
$z_j=q^2z_k(j< k)$, $u_j=q^2u_k(j< k)$.
$\Box$

Now let us decompose $<v^\prime|\ozu|v>$ in the following manner:
\begin{eqnarray}
&&
<v^\prime|\ozu|v>=
\sum_{j=1}^{n-1}{\ojzu\over u_j-q^2u_{j+1}}
+\otzu,
\label{decomp}
\\
&&
\ojzu=\hbox{Res}_{u_j=q^2u_{j+1}}
(<v^\prime|\ozu|v>-\sum_{k=1}^{j-1}{\okzu\over u_k-q^2u_{k+1}})
\hbox{ for $j\geq 2$},
\nonumber
\\
&&
\oizu=
\hbox{Res}_{u_1=q^2u_2}
<v^\prime|\ozu|v>.
\nonumber
\end{eqnarray}
Then

\begin{lemma}
\begin{description}
\item[(1)] $\ojzu\in \sum N_k\ot (V_1)_{z_1}\ot\cdots\ot(V_1)_{z_{n+1}}$,
\item[(2)] $\otzu$ is regular at $u_j=q^{2(k-j)}u_k(j<k)$,
\item[(3)] $\ojzu$ is regular at $u_r=qz_r(1\leq r\leq n)$,
\item[(4)] $\ojzu|_{u_r=qz_r(1\leq r\leq n)}=0$.
\end{description}
\end{lemma}
\par
\noindent
Proof:(1): This follows from \ref{typeii}.\par
\noindent (3): This is obvious from proposition \ref{regular} (2).
\par
\noindent
(4): It follows from
\begin{eqnarray}
&&
{1\over f}|_{u_l=q^2u_{l+1}}=g^{-1}
\prod_{j<k}{({q^4z_k\over z_j})_\infty \over({q^2z_k\over z_j})_\infty}
\prod_{j>k,j,k\neq l,l+1}
{({q^2u_k\over u_j})_\infty \over({u_k\over u_j})_\infty}
\prod_{j\neq l,l+1}\prod_{k}
{({u_j\over qz_k})_\infty \over({qu_j\over z_k})_\infty}
{\prod_{k=1}^{n+1}(1-{u_{l+1}\over qz_k})\over
\prod_{j=l+2}^n(1-{u_{l+1}\over u_j})}
\nonumber
\end{eqnarray}
and \ref{typeii} that
$Res_{u_l=q^2u_{l+1}}<v^\prime|\ozu|v>$ has
$\prod_{k=1}^{n+1}(1-{u_{l+1}\over qz_k})$
as a factor of its zero divisor.
Taking further residues does not produce poles at
$u_{l+1}=qz_k(1\leq k\leq n)$ by proposition \ref{regular} (2).
Hence $\ojzu|_{u_r=qz_r(1\leq r\leq n)}=0$.
\par
\noindent
(2): Let us prove, for $2\leq j\leq n$, that
\begin{eqnarray}
&&
<v^\prime|\ozu|v>-
\sum_{r=1}^{j-1}{\orzu\over u_r-q^2u_{r+1}}
\hbox{ is regular at $u_l=q^{2(s-l)}u_s(l<s,1\leq l\leq j-1)$}
\nonumber
\end{eqnarray}
by the induction on $j$.
The $j=2$ case is obvious from proposition \ref{regular} (2).

Suppose that the statement is true for $1\leq j\leq k$.
We have
\begin{eqnarray}
&&
<v^\prime|\ozu|v>-
\sum_{r=1}^{k}{\orzu\over u_r-q^2u_{r+1}}
=O^{(1)}({\bf z}|{\bf u})-\okzu,
\nonumber
\\
&&
O^{(1)}({\bf z}|{\bf u})
=<v^\prime|\ozu|v>-
\sum_{r=1}^{k-1}{\orzu\over u_r-q^2u_{r+1}},
\nonumber
\\
&&
\okzu=Res_{u_k=q^2u_{k+1}}O^{(1)}({\bf z}|{\bf u}).
\nonumber
\end{eqnarray}
By the induction hypothesis $O^{(1)}({\bf z}|{\bf u})$ is regular
at $u_l=q^{2(s-l)}u_s(l<s,1\leq l\leq k-1)$.
Hence $\okzu$ and consequently
$O^{(1)}({\bf z}|{\bf u})-\okzu$ are regular
at $u_l=q^{2(s-l)}u_s(l<s,1\leq l\leq k-1)$.
The definition of a residue and proposition \ref{regular} (2)
imply that $O^{(1)}({\bf z}|{\bf u})-\okzu$ is
regular at $u_k=q^{2(s-k)}u_s(k<s)$.
Hence the statement is proved for $j=k+1$.
$\Box$
\vskip2truemm

Using the decomposition \ref{decomp} we have
\begin{eqnarray}
Pr(u)e_0A^\prime\ot B=&&
\sum_{j=1}^{n-1}
{1\over u_j-q^2u_{j+1}}Pr(u)Pr(z)
(e_0\ot1)\ojzu
\nonumber
\\
&&
+
Pr(u)Pr(z)(e_0\ot1)(1-Pr(u))\otzu.
\nonumber
\end{eqnarray}
Note that, in $(V_1)_{u_1}\ot(V_1)_{u_2}$
\begin{eqnarray}
&&
e_0w=(u_1-q^2u_2)v^{(1)}_1\ot v^{(1)}_1.
\nonumber
\end{eqnarray}
Since $\otzu$ has no poles at $u_j=q^{2(s-j)}u_s(j<s)$
we can conclude that
\begin{eqnarray}
&&
Pr(u)Pr(z)(e_0\ot1)(1-Pr(u))\otzu|_{u_j=q^{-2(j-1)+1}z}=0.
\nonumber
\end{eqnarray}
Since each $\ojzu$ has a zero divisor of the form
$\prod_{j=1}^{n+1}(1-{u_l\over qz_j})$
\begin{eqnarray}
&&
\sum_{j=1}^{n-1}
{1\over u_j-q^2u_{j+1}}Pr(u)Pr(z)
(e_0\ot1)\ojzu|_{u_j=qz_j(1\leq j\leq n)}=0.
\nonumber
\end{eqnarray}
Taking into account that $Pr(u)e_0A^\prime\ot B$ has no pole
at all we can conclude that
\begin{eqnarray}
&&
Pr(u)e_0A^\prime\ot B
|_{z_j=q^{-2(j-1)}z(1\leq j\leq n+1),u_j=q^{-2(j-1)}z(1\leq j\leq n)}
=0.
\nonumber
\end{eqnarray}
Hence \ref{eqi} is proved.
The equation \ref{eqiii} is similarly proved.
$\Box$
\vskip5mm

\begin{cor}
\begin{eqnarray}
\phiunnp(z)&=&
(-1)^{[{n\over2}](n-1)}
(-q)^{-{n(n-2)\over4}+{3\over8}(1-(-1)^n)}O(q^nz)
\hbox{ on $V(\La_0)$}
\nonumber
\\
&=&
(-1)^{[{n\over2}]}
(-q)^{-{n\over12}(2n^2-9n+10)-{3\over8}(1-(-1)^n)}
z^{{n(n-1)\over2}}O(q^nz)
\hbox{ on $V(\La_1)$}.
\nonumber
\end{eqnarray}
\end{cor}

\section{Commutation relations of vertex operators}
Using the fusion construction we can determine the
commutation relations of vertex operators in section \ref{EVO} and
\ref{FVO}.
Here we give only commutation relations
which is used for the application
to the vertex models.

\begin{theor}\label{thcr}
\begin{description}
\item[(1)]
\begin{eqnarray}
&&
\udphinnp(z)\phinnp(z)=c_i\id_{(\Vn)_z\ot \Vli},
\nonumber
\\
&&
c_i=(-1)^{[{n\over2}]n+{1\over6}n(n-1)(n+4)}
q^{-{1\over6}n(n^2-7)+in(n-2)}
\nonumber
\\
&&
\quad\quad\quad
[n+1]^{1-i}z^{{n(n-1)\over2}}
{(q^2)_\infty(q^4)_\infty\over (q^{2n+2})_\infty^2(q^2;q^2)_n},
\nonumber
\end{eqnarray}
where $(z;p)_n=\prod_{l=0}^{n-1}(1-zp^l)$.
\item[(2)]
\begin{eqnarray}
&&
(-1)^nq^{-3[{n+1-i\over2}]+n(i+1)}\big({z\over w}\big)^{i-{1\over2}}
\check{R}_{n+1,1}({z\over w})O(q^nz)\Phi(w)=\Phi(w)O(q^nz),
\nonumber
\\
&&
\check{R}_{n+1,1}(z)=z^{-{1\over2}}
r_{n+1}(z)\brc_{n+1,1}(z),
\quad
r_{n+1}(z)=
{(q^{n+2}z)_\infty(q^{n+4}z^{-1})_\infty
\over
(q^{n+2}z^{-1})_\infty(q^{n+4}z)_\infty}
\nonumber
\end{eqnarray}
\end{description}
\end{theor}

\begin{prop}\label{propo}
\begin{eqnarray}
&&
P_F^{n+1}O(q^{-2}z)\oz=
(-1)^{n+i}q^{{n(n+1)\over2}+i}
{
(q^2)_\infty(q^4)_\infty
\over
(q^{2n+2})^2_{\infty} (q^2;q^2)_n
}
w_n\ot\id_{\Vli},
\nonumber
\end{eqnarray}
where $w_n$ is the highest weight vector with weight zero
in $\Vn\ot\Vn$ which is explicitly described in \ref{hwti}
and \ref{hwtii}.
\end{prop}

We shall first prove Proposition \ref{propo} and after that
deduce (1) of Theorem \ref{thcr}.
Let us set
\begin{eqnarray}
&&
\otzzuu=
\nonumber
\\
&&
\phiad(z_1^\prime)\Phi(z_1)
\cdots
\phiad(z_{n+1}^\prime)\Phi(z_{n+1})
\Psi(q^{-2}u_n^\prime)\psiamd(q^{-2}u_n)
\cdots
\Psi(q^{-2}u_1^\prime)\psiamd(q^{-2}u_1).
\nonumber
\end{eqnarray}
Using the commutation relations of the vertex operators
$\Phi(z)$ and $\Psi(z)$, we have
\begin{eqnarray}
&&
{h\over ff^\prime}(-1)^{{n(n-1)\over2}+i}q^{n+i}
\prod_{j<k}\big({z_j\over z_k^\prime}\big)^{\overline{i+j-k}}
\prod_{j<k}\big({u_j^\prime\over u_k}\big)^{\overline{i+j-k-1}}
\prod_{j,k}\big({u_j^\prime\over q^2z_k}\big)^{\overline{i+j-k-1}}
\nonumber
\\
&&
\brc({u_1^\prime\over q^2u_n})
\cdots
\brc({u_1^\prime\over q^2u_2})
\cdots
\brc({u_{n-1}^\prime\over q^2u_n})
\brc({q^2z_1\over z_{n+1}^\prime})
\cdots
\brc({q^2z_1\over z_2^\prime})
\cdots
\brc({q^2z_n\over z_{n+1}^\prime})
\otzzuu
\nonumber
\\
&&
=\oqtzpup\ozu,
\label{eq1}
\\
&&
g=\prod_{j<k}r({q^2z_j\over z_k^\prime})
\prod_{j<k}r({u_j^\prime\over q^2u_k})
\prod_{j,k}
{
\theta_{q^4}({q^3z_j\over u_k^\prime})
\over
\theta_{q^4}({u_k^\prime\over qz_j})
}
\nonumber
\end{eqnarray}
where $f^\prime=f({\bf z}^\prime|{\bf u}^\prime)$,
$q^{-2}{\bf z}^\prime=
(q^{-2}z_1^\prime,\cdots,q^{-2}z_{n+1}^\prime)$ etc.
 and
$\overline{i+j-k}$ etc. means that the number $i+j-k$ should
be considered modulo two, inparticular
$\overline{i+j-k}=0\hbox{ or }1$.
Note that
\begin{eqnarray}
&&
h|_{u_j=q^3z_{j+1},u_j^\prime=q^3z_{j+1}^\prime(1\leq j\leq n)}
=\prod_{j=2}^{n+1}(1-{z_j\over z_j^\prime})\tilde{h},
\nonumber
\\
&&
\tilde{h}=
q^{-n(n+1)}
\prod_{2\leq j<k\leq n+1}
{
({q^4z_j\over z_k^\prime})_\infty^2
({q^4z_j^\prime\over z_k})_\infty
({z_j^\prime\over z_k})_\infty
(1-{z_k\over z_j^\prime})
\over
({q^2z_j\over z_k^\prime})_\infty
({q^6z_j\over z_k^\prime})_\infty
({z_j^\prime\over q^2z_k})_\infty
({q^2z_j^\prime\over z_k})_\infty
}
\prod_{j=2}^{n+1}
{
({q^4z_1\over z_j^\prime})_\infty^2
({q^4z_j\over z_j^\prime})_\infty
({q^4z_j^\prime\over z_j})_\infty
\over
({q^6z_1\over z_j^\prime})_\infty
({q^2z_1\over z_j^\prime})_\infty
({q^2z_j^\prime\over z_j})_\infty
({q^2z_j\over z_j^\prime})_\infty
}.
\nonumber
\end{eqnarray}
Specializing the variables to
$u_j=q^3z_{j+1},u_j^\prime=q^3z_{j+1}^\prime(1\leq j\leq n)$
in both hand sides of the equation \ref{eq1}, after that
setting $z_j=z_j^\prime (1\leq j\leq n+1)$ and
using \ref{invii}
\begin{eqnarray}
&&
\lim_{z_j\rightarrow z_j^\prime}
(1-{z_j\over z_j^\prime})
\Psi(z_j^\prime) \psiamd(z_j)
=-gw\ot\id_{\Vli}
\nonumber
\end{eqnarray}
we have
\begin{eqnarray}
&&
{\tilde{\tilde{h}}\over \tilde{f}^2}
(-1)^{{n(n+1)\over2}+i}q^{n+i}
g^n
\prod_{1\leq j<k\leq n+1}\big({z_j\over z_k}\big)^{\overline{i+j-k}}
\prod_{2\leq j<k\leq n+1}\big({z_j\over z_k}\big)^{\overline{i+j-k-1}}
\prod_{j=2}^{n+1}\prod_{k=1}^{n+1}
\big({qz_j\over z_k}\big)^{\overline{i+j-k}}
\nonumber
\\
&&
\prod_{2\leq j<k\leq n+1}
{1\over 1-{z_j\over q^2z_k}}
R_n(z)w^{\ot n}
\ot
\tilde{R}_{n+1}(z)
\phiad(z_1)\Phi(z_1)
\cdots
\phiad(z_{n+1})\Phi(z_{n+1})
\nonumber
\\
&&
O(q^{-2}{\bf z}|qz_{n+1},\cdots,qz_2)
O({\bf z}|q^3z_{n+1},\cdots,q^3z_2),
\label{fundeq}
\end{eqnarray}
where
\begin{eqnarray}
&&
R_n({\bf z})=
\brc({z_2\over q^2z_{n+1}})
\cdots
\brc({z_2\over q^2z_3})
\cdots
\brc({z_n\over q^2z_{n+1}})
\nonumber
\\
&&
\tilde{R}_{n+1}({\bf z})
=
\brc({q^2z_1\over z_{n+1}})
\cdots
\brc({q^2z_1\over z_2})
\cdots
\brc({q^2z_n\over z_{n+1}}),
\nonumber
\\
&&
\tilde{\tilde{h}}
=q^{-n(n+1)}g^{-2n}
\prod_{2\leq j<k\leq n+1}
{
({q^4z_j\over z_k})_\infty^3
({z_j\over z_k})_\infty
(1-{z_k\over z_j})
\over
({q^2z_j\over z_k})_\infty^3
({q^6z_j\over z_k})_\infty
}
\prod_{j=2}^{n+1}
{
({q^4z_1\over z_j})_\infty^2
\over
({q^6z_1\over z_j})_\infty
({q^2z_1\over z_j})_\infty
},
\nonumber
\\
&&
\tilde{f}
=g^{-n}
\prod_{2\leq j<k\leq n+1}
{
({z_j\over z_k})_\infty
({q^4z_j\over z_k})_\infty
\over
({q^2z_j\over z_k})_\infty^2
},
\nonumber
\end{eqnarray}
and $w=v^{(1)}_0\ot v^{(1)}_1-qv^{(1)}_1\ot v^{(1)}_0$.

\begin{lemma}\label{pri}
Let $Pr_n$ be the $\uqi$ linear projection
$V_1^{\ot n}\ot V_1^{\ot n}\ra\Vn\ot\Vn$. Then we have
\begin{eqnarray}
&&
Pr_nR_n({\bf z})w^{\ot n}=
q^{{n(n-1)\over2}}
\prod_{2\leq j<k\leq n+1}
{
1-{z_j\over q^2z_k}
\over
1-{z_j\over z_k}
}
w_n.
\nonumber
\end{eqnarray}
\end{lemma}
\par
\noindent
Proof: Since $Pr_nR_n({\bf z})w^{\ot n}$ belongs to
the trivial representation of $\Vn\ot\Vn$, we have
$Pr_nR_n({\bf z})w^{\ot n}=cw_n$ for some scalar function $c$.
The function $c$ is the coefficient of $v^{(n)}_0\ot v^{(n)}_n$
in the right hand side.
Let us calculate the coefficient of $v^{(1)\ot n}_0\ot v^{(1)\ot n}_1$
in $R_n({\bf z})w^{\ot n}$. It is easy to see that this coefficient
is the same as that of $v^{(1)\ot n}_0\ot v^{(1)\ot n}_1$
in $R_n({\bf z})(v^{(1)}_0\ot v^{(1)}_1)^{\ot n}$.
The latter coefficient is easily calculated and coinsides with
the function in the statement of the lemma.
$\Box$
\vskip1truecm

Let
$(P^1_F)^{\ot (n+1)}$
be the $\Up$ linear map
$(V_1)_{z_1}^{a\ast}\ot (V_1)_{z_1}\ot
\cdots \ot(V_1)_{z_{n+1}}^{a\ast}\ot (V_1)_{z_{n+1}}\ra F$
defined by
$(P^1_F)^{\ot (n+1)}
\big(
\ot_{l=1^{n+1}}
(
v^{(1)\ast}_{j_l}\ot v^{(1)}_{k_l}
)
\big)
=\prod_{l=1}^{n+1}\delta_{j_l,k_l}$
and $P_F^{n+1}$ the dual pairing map
$(V_{n+1})_{q^{-n}z}^{\ast a}\ot
(V_{n+1})_{q^{-n}z}\ra F$.

\begin{lemma}\label{scalar}
There is an equation
\begin{eqnarray}
&&
c P^{n+1}_F Pr_{n+1}\tilde{R}_{n+1}({\bf z})=(P^1_F)^{\ot (n+1)},
\nonumber
\\
&&
c=q^{-{n(n+1)\over2}}
{
(q^2;q^2)_{n+1}
\over
(1-q^{2})^{n+1}
}
\nonumber
\end{eqnarray}
at $z_j=q^{-2(j-1)}z (1\leq j\leq n+1)$.
\end{lemma}
\par
\noindent
Note that the R-matrix $\br({q^2z_j\over z_k})(j<k)$
is regular at ${z_j\over z_k}=q^{2(k-j)}$
and $\br({q^2z_j\over z_k})^{-1}=\br({z_k\over q^2z_j})$
which is also regular at ${z_j\over z_k}=q^{2(k-j)}$.
Hence there exists the inverse of $\tilde{R}_{n+1}({\bf z})$
which is regular at $z_j=q^{-2(j-1)} (1\leq j\leq n+1)$.
Let us set
$\varphi({\bf z})=(P^1_F)^{\ot (n+1)}\tilde{R}_{n+1}^{-1}({\bf z})$,
\[
\begin{array}{ccc}
(V_1)_{z_1}^{\ast a}\ot(V_1)_{z_1}
\ot\cdots\ot
(V_1)_{z_{n+1}}^{\ast a}\ot(V_1)_{z_{n+1}}
&
\stackrel{(P^1_F)^{\ot (n+1)}}{\ra}
&
F
\\
\downarrow\tilde{R}_{n+1}({\bf z})
&
{}
&
\downarrow\id
\\
(V_1)_{z_1}^{\ast a}\ot\cdots(V_1)_{z_{n+1}}^{\ast a}
\ot(V_1)_{z_1}\ot\cdots\ot(V_1)_{z_{n+1}}
&
\stackrel{\varphi({\bf z})}{\ra}
&
F.
\end{array}
\]
Then

\newtheorem{sub}{sublemma}

\begin{sub}
\begin{eqnarray}
&&
\varphi({\bf z})(N_j\ot V_{z}\ot \cdots \ot V_{q^{-2n}z})
=
\varphi({\bf z})(V_{z}^{\ast a}\ot \cdots \ot V_{q^{-2n}z}^{\ast a}\ot N_j)
=0
\nonumber
\end{eqnarray}
for all $1\leq j\leq n+1$.
\end{sub}
\vskip2truemm
\noindent
Proof: Since $\varphi({\bf z})$ is $\Up$ linear map we have
\begin{eqnarray}
&&
\varphi({\bf z})(v^{(1)\ast}_{j_1}\ot\cdots\ot v^{(1)\ast}_{j_{n+1}}\ot
v^{(1)}_{k_1}\ot\cdots \ot v^{(1)}_{k_{n+1}})
\nonumber
\\
&&
=\beta<v^{(1)\ast}_{j_{n+1}}\ot\cdots\ot v^{(1)\ast}_{j_1},
\ttr_{n+1}(z)(v^{(1)}_{k_1}\ot\cdots \ot v^{(1)}_{k_{n+1}})>,
\label{ttr}
\end{eqnarray}
for some scalar function $\beta$.
Here $\ttr_{n+1}(z)$ is the specialization
of the variables to $z_j=q^{-2(j-1)}z(1\leq j\leq n+1)$
of the $\Up$ intertwiner
$(V_1)_{z_1}\ot\cdots\ot(V_1)_{z_{n+1}}
\ra (V_1)_{z_{n+1}}\ot\cdots\ot(V_1)_{z_1}$
normalized as
$\ttr_{n+1}({\bf z})(v^{(1)\ot n}_0)=v^{(1)\ot n}_0$.
In fact, for generic values of $z_j's$ for which
$(V_1)_{z_1}^{\ast a}\ot\cdots\ot(V_1)_{z_{n+1}}^{\ast a}
\ot (V_1)_{z_1}\ot\cdots\ot(V_1)_{z_{n+1}}$
is irreducible,
the $\Up$ linear map
$(V_1)_{z_1}^{\ast a}\ot\cdots\ot(V_1)_{z_{n+1}}^{\ast a}
\ot (V_1)_{z_1}\ot\cdots\ot(V_1)_{z_{n+1}}\ra F$
is unique up to scalar factor
and given by $\ttr_{n+1}({\bf z})$ as in the right hand side of \ref{ttr}.
Since
$\beta=\varphi({\bf z})
(v^{(1)\ast\ot(n+1)}_0\ot v^{(1)\ot(n+1)}_0)$
and
$\tr_{n+1}({\bf z})$
is regular at
$z_j=q^{2(k-j)}z_k(j<k)$,
$\beta$ is also regular at
$z_j=q^{2(k-j)}z_k(j<k)$.
Hence \ref{ttr} holds at
$z_j=q^{-2(j-1)}z(1\leq j\leq n+1)$.
By Proposition \ref{ImR} we have
$\ttr_{n+1}({\bf z})(N_j)=0$
and hence
$\varphi({\bf z})(V_{z}^{\ast a}\ot \cdots\ot
V_{q^{-2n}z}^{\ast a}\ot N_j)=0$.

Let us prove the remaining equation.
Note that the base of the trivial representation in
$V_{u}^{\ast a}\ot V_{q^{-2}u}^{\ast a}$ is given by
$v^{(1)\ast}_1\ot v^{(1)\ast}_0-qv^{(1)\ast}_0\ot v^{(1)\ast}_1$.
Taking into account the fact that,
in the left part of the right hand side of
the equality \ref{ttr}, the order of the tensor product
is reversed, we set
$w^\ast=v^{(1)\ast}_0\ot v^{(1)\ast}_1-qv^{(1)\ast}_1\ot v^{(1)\ast}_0$.
Then, by calculations, we have
\begin{eqnarray}
&&
<w^{\ast},f_1^k(v^{(1)}_0\ot v^{(1)}_0)>=0
\hbox{ for $0\leq k\leq 2$}.
\nonumber
\end{eqnarray}
Since, by Proposition \ref{ImR},
$Im\ttr_{n+1}({\bf z})\simeq (V_{n+1})_{q^{-n}z}$
which is generated by $v^{(1)\ot(n+1)}_0$ over $\uqi$,
we have
\begin{eqnarray}
&&
\varphi({\bf z})(N_j\ot V_{z}\ot \cdots\ot V_{q^{-2n}z})=0.
\nonumber
\end{eqnarray}
$\Box$
\vskip5truemm

Let us continue the proof of lemma.
By the sublemma the map $\varphi({\bf z})$ induces
the $\Up$ linear map
\begin{eqnarray}
&&
(V_{n+1})_{q^{-n}z}^{\ast a}\ot
(V_{n+1})_{q^{-n}z}
\ra F
\nonumber
\end{eqnarray}
which we denote the same symbol.
Hence $\varphi({\bf z})$ is a scalar multiple of the
canonical pairing map $P^{n+1}_F$, $\varphi({\bf z})=cP_F^{n+1}$.

Let us determine the scalar $c$.
Note that
$c=\varphi({\bf z})
((v^{(1)}_0)^{\ot (n+1)}\ot (v^{(1)}_1)^{\ot (n+1)})$.
It is easily proved that
\begin{eqnarray}
&&
\varphi({\bf z})
((v^{(1)}_0)^{\ot (n+1)}\ot (v^{(1)}_1)^{\ot (n+1)})
=<(v^{(1)\ast}_0\ot v^{(1)\ast}_1)^{\ot (n+1)},
\tr_{n+1}^{-1}(z)(v^{(1)}_0)^{\ot (n+1)}\ot (v^{(1)}_1)^{\ot (n+1)})>.
\nonumber
\end{eqnarray}
Recall that
\begin{eqnarray}
&&
\tr_{n+1}^{-1}(z)=
\brc({z_{n+1}\over q^2z_n})\cdots
\brc({z_2\over q^2z_1})
\cdots
\brc({z_{n+1}\over q^2z_1})
\nonumber
\end{eqnarray}
with $z_j=q^{-2(j-1)}z (1\leq j\leq n+1)$.

It follows from those description we have
\begin{eqnarray}
&&
c=q^{{n(n+1)\over2}}
\prod_{1\leq j<k\leq n+1}
{1-{z_k\over q^2z_j}\over1-{z_k\over z_j}}
=
q^{{n(n+1)\over2}}
{
\prod_{l=1}^{n+1}(1-q^{-2l})
\over
(1-q^{-2})^{n+1}
}.
\nonumber
\end{eqnarray}
$\Box$
\vskip1truecm

\noindent
Proof of Proposition \ref{propo}:
Taking $(1\ot P^{n+1}_F)(Pr_n\ot Pr_{n+1})$ in both hand sides
of the equation \ref{fundeq} and
using Lemma \ref{pri}, Lemma \ref{scalar}, equation \ref{invi},
 we have the equality in the statement of
Proposition \ref{propo}.
$\Box$
\vskip1truecm

Now (1) of Theorem \ref{thcr} is derived from
Proposition \ref{propo} in the following manner.
Let us introduce the $\Up$ intertwiners
\begin{eqnarray}
&&
\phinnpad(z):
\Vli\ra(\Vn)_z\ot\Vlip\ot(\Vnp)_z^{\ast a},
\nonumber
\\
&&
\phinamdnp(z):
\Vli\ra(\Vn)^{\ast a^{-1}}_z\ot\Vlip\ot(\Vnp)_z,
\nonumber
\end{eqnarray}
by
\begin{eqnarray}
&&
<\vnpj,\phinnpad(z)u>=
\udphinnp(z)(u\ot\vnpj),
\nonumber
\\
&&
<\vnj,\phinamdnp(z)u>=
\phinnp(z)(\vnj\ot u).
\nonumber
\end{eqnarray}
Then
\begin{eqnarray}
&&
\phinnpad(z)=(-1)^{n+1-i}q^{-(n+1)(1-i)}[n+1]^i\phiunnp(q^{-2}z),
\nonumber
\\
&&
\phinamdnp(z)=\phiunnp(z).
\nonumber
\end{eqnarray}
Those equations implies that the equation
\begin{eqnarray}
&&
\udphinnp(z)\phinnp(z)=\gamma\id_{(\Vn)_z\ot \Vli}
\nonumber
\end{eqnarray}
is equivalent to
\begin{eqnarray}
&&
P^{n+1}_F\phiunnp(q^{-2}z)\phiunnp(z)=
(-1)^{n+i}q^{i(n+1)}[n+1]^{i-1}\gamma
\sum_{j=0}^n\vdnj\ot\vnj\ot\id_{\Vli}.
\nonumber
\end{eqnarray}
Rewriting these equations in terms of $\oz$, we have
\begin{eqnarray}
P_F^{n+1}O(q^{n-2}z)O(q^nz)
&=&\gamma
(-1)^{[{n\over2}]n+n+i+{1\over6}n(n-1)(n+4)}
q^{{1\over6}n(n-1)(n+4)+i(-n^2+2n+1)}
\nonumber
\\
&&
z^{-{n(n-1)\over2}}[n+1]^{-1+i}
\big(\sum_{j=0}^n\vdnj\ot\vnj\big)
\ot\id_{\Vli}.
\nonumber
\end{eqnarray}
Theorem \ref{thcr} (1) follows from
this equation and Proposition \ref{propo}.
$\Box$
\vskip1.5truecm

\noindent
Proof of Theorem \ref{thcr} (2):
Using the commutaion relations of $\Phi(z)$ and $\Psi(z)$
we have
\begin{eqnarray}
&&
(-1)^{n+1}
\prod_{l=1}^n\big({u_l\over w}\big)^{-\overline{i+l}}
\prod_{j=1}^{n+1}\big({z_j\over w}\big)^{\overline{i+l-1}}
\prod_{j=1}^{n+1}r({z_j\over w})
\prod_{l=1}^{n}
{
\theta_{q^4}({qu_l\over w})
\over
\theta_{q^4}({qw\over u_l})
}
\brc({z_1\over w})
\cdots
\brc({z_{n+1}\over w})
\ozu\Phi(w)
\nonumber
\\
&&
=\Phi(w)\ozu.
\label{symni}
\end{eqnarray}
Similarly to the Proposition \ref{regular}, we can prove that
$(Pr(z)\ot Pr(u))\ozu\Phi(w)$ and
$(Pr(z)\ot Pr(u))\Phi(w)\ozu$ give well-difined
$\Up$-intertwinwers at $z_j=q^{-2j+2}z$,
$u_j=q^{-2j+3}z$.
Hence, by Theorem \ref{vo}, we have
\begin{eqnarray}
&&
(Pr(z)\ot Pr(u))
\Big[
\brc({z_1\over w})
\cdots
\brc({z_{n+1}\over w})
\ozu\Phi(w)
\Big]_{z_j=q^{-2j+2}z,u_j=q^{-2j+3}z}
\\
&&
=c(z,w)
\brc_{n+1,1}({q^{-n}z\over w})
\Big[
(Pr(z)\ot Pr(u))
\ozu\Phi(w)
\Big]_{z_j=q^{-2j+2}z,u_j=q^{-2j+3}z}
\end{eqnarray}
for some scalar function $c(z,w)$.
Comparing the coefiicient of $v^{(1)}_0\ot v^{(n)}_0$
we conclude that $c(z,w)\equiv1$.
Taking $Pr(z)\ot Pr(u)$ of the both hand sides of the
equation \ref{symni} and
substituting $z_j=q^{-2(j-1)}z (1\leq j\leq n+1)$,
$u_j=q^{-2j+3}z (1\leq j\leq n)$,
we obtain the desired equation.
$\Box$
\vskip2truecm

\section{Inhomogeneous vertex model of $6$-vertex type}
In this section we denote $(V_s)_1$ by $V_s$ for the sake of simplicity
and assume $-1<q<0$.
Let us consider the two dimensional regular square infinite lattice.
Fix the positive integer $N$ and
non-negative integers $s_1,\cdots,s_N$ and
vertical lines $l_1,\cdots,l_N$.
Then the vertex model which we study here is defined in the following way.
We associate the representation $V_1$ of $\Up$ with each edge on
horizontal lines and on vertical lines except $l_1,\cdots,l_N$.
With each edge on the line $l_j$ we associate the vector space $V_{s_j}$.
For each vertex the Boltzmann weight is given by the corresponding
R-matrix, $R_{11}(1)$, $R_{s_j1}(1)$.
We can assume that the lines $l_1,\cdots,l_N$ are successive
by including $1$ in the set of $s_j$.
Let us first give the mathematical objects and after that
explain the validity of them.
We use the following vertex operators in this section.
\begin{defi}
The intertwiners
\begin{eqnarray}
&&
\ouunnp(z):
\Vli\ra (\Vn)_{q^2z}\ot \Vlip\ot (\Vnp)_z,
\nonumber
\\
&&
\odunnp(z):
(\Vn)_z\ot V(\La_i)\ra \Vlip\ot (\Vnp)_z,
\nonumber
\\
&&
\oudnnp(z):
\Vli\ot (\Vnp)_z\ra (\Vn)_z\ot \Vlip,
\nonumber
\end{eqnarray}
are defined by
\begin{eqnarray}
&&
\ouunnp(z)=O(q^nz),
\quad
\odunnp(z)(\vnj\ot\cdot)=<\vnj,\ouunnp(z)>,
\nonumber
\\
&&
\oudnnp(z)(\cdot\ot\vnpj)=<\vnpj,\ouunnp(q^{-2}z)>,
\nonumber
\end{eqnarray}
where the pairing is defined by the isomorphism
$(V_n)_{q^2z}\simeq(V_n)_{z}^{\ast a^{-1}}$,
$(V_{n+1})_{q^{-2}z}\simeq(V_{n+1})_{z}^{\ast a}$.
\end{defi}

Theorem \ref{thcr} (2) and Proposition \ref{propo}
imply the commutation relations
\begin{eqnarray}
&&
(-1)^nq^{3[{n+1-i\over2}]-n(i+1)}
\big({w\over z}\big)^{i-{1\over2}}
\check{R}_{1,n+1}({w\over z})
\Phi(w)\odunnp(z)=\odunnp(z)\Phi(w)
\label{ocom}
\\
&&
\oudnnp(z)\odunnp(z)=
(-1)^{n+i}q^{{n(n+1)\over2}+i}
{(q^2)_\infty(q^4)_\infty\over
(q^{2n+2})^2_\infty(q^2:q^2)_n}
\id_{(V_n)_z\ot\Vli}.
\nonumber
\end{eqnarray}
on $\Vli$.
The representation theoretical formulation for the model is given by

\begin{description}
\item[({\bf Space})]
The space acted by the transfer matrix is
\begin{eqnarray}
&&\H=\oplus_{i,j=0,1}\H_{s_N\cdots s_1,ij},
\nonumber
\\
&&
\H_{s_N\cdots s_1,ij}=V_{s_N-1}\ot\cdots\ot V_{s_1-1}\ot
\Vli\ot\Vlj^{\ast a}
\nonumber
\end{eqnarray}
\item[({\bf Transfer matrix})]
The transfer matrix is given by
\begin{eqnarray}
&&
T(z)=\hbox{id}\ot T_{XXZ}(z),
\nonumber
\end{eqnarray}
where $T_{XXZ}(z)$ is the transfer matrix of the
6-vertex model actiong on $\oplus_{i,j=0,1}\Vli\ot\Vlj^{\ast a}$.
Explicitly $T_{XXZ}(z)=g\Phi^\ast(z)\Phi(z)$ and
$\Phi^\ast(z)={}^t\Phi(q^{2}z):
(V_1)_{z}\ot\Vli^{\ast a}\ra\Vlip^{\ast a}$, where
$g={(q^2)_\infty\over(q^4)_\infty}$.
\item[({\bf Ground state})]
The space of vacuum vectors $V_{vac}$ is
\begin{eqnarray}
&&
V_{vac}=
\oplus_{i,j=0,1}V_{s_N-1}\ot \cdots \ot V_{s_1-1}\ot F|vac>_{XXZ,i},
\nonumber
\end{eqnarray}
where $|vac>_{XXZ,i}$ is the vacuum vector
of the XXZ-model in $\Vli\ot\Vli^{\ast a}$.
\item[({\bf Excited states})]
The creation and annihilation operators are given by
\begin{eqnarray}
&&
\varphi^\ast_j(z)=1\ot\varphi^\ast_{j,XXZ}(z),
\quad
\varphi_j(z)=1\ot\varphi_{j,XXZ}(z),
\nonumber
\end{eqnarray}
where $\varphi^\ast_{j,XXZ}(z)$, $\varphi_{j,XXZ}(z)$
are the creation and annihilation operators of the
XXZ model,
\begin{eqnarray}
&&
\varphi^\ast_{j,XXZ}(z)=<\vij,\psiamd(z)>,
\quad
\varphi_{j,XXZ}(z)=<\vdij,\Psi(z)>.
\nonumber
\end{eqnarray}
\item[({\bf Local operators})]
For $L\in\hbox{End}(V_{s_N}\ot \cdots \ot V_{s_1})$ the corresponding
local operator ${\cal L}$ is defined by
\begin{eqnarray}
&&
{\cal L}=\Phi^{s_N,\cdots,s_1}(1,\cdots,1)^{-1}(1\ot L)
\Phi^{s_N,\cdots,s_1}(1,\cdots,1),
\nonumber
\\
&&
\Phi^{s_N,\cdots,s_1}(z_N,\cdots,z_1)={}_{s_N-1}O^{s_N}(z_N)
\cdots{}_{s_1-1}O^{s_1}(z_1),
\nonumber
\\
&&
\Phi^{s_N,\cdots,s_1}(z_N,,\cdots,z_1)^{-1}=c_{i,N}
{}^{s_1-1}O_{s_1}(z_1)\cdots{}^{s_N-1}O_{s_N}(z_N)
\hbox{ on $\H_{s_N\cdots s_1,ij}$},
\nonumber
\\
&&
c_{i,N}=
(-1)^{\sum_{j=1}^Ns_j+iN+{N(N-3)\over2}}
q^{-\sum_{j=1}^N{s_j(s_j-1)\over2}-\sum_{j=0}^{N-1}\overline{i+j}}
{
\prod_{j=1}^N(q^{2s_j})^2_\infty(q^2;q^2)_{s_j-1}
\over
(q^2)_\infty^N(q^4)_\infty^N
}.
\nonumber
\end{eqnarray}

\item[({\bf Correlation functions})]
The expectation values of the local operator ${\cal L}$
is given by
\begin{eqnarray}
&&
<{\cal L}>_i=
{
\hbox{tr}_{V_{s_N-1}\ot \cdots \ot V_{s_1-1}\ot\Vli}
((1\ot q^{-2\rho}){\cal L})
\over
s_1\cdots s_N\hbox{tr}_{\Vli}(q^{-2\rho})
},
\nonumber
\end{eqnarray}
where $\rho=\La_0+\La_1$ and $1$ is the identity operator
acting on $V_{s_N-1}\ot \cdots \ot V_{s_1-1}$.

\end{description}

Let us explain why we have given the mathematical setting
as above.
The less obvious definition is that of the transfer matrix.
If it is permitted then others are rather natural
compared with the case of the XXZ model.
So we shall explain the reason of our definition
of the transfer matrix.
The natural definition of the transfer matrix $T(z)$ should be
\begin{eqnarray}
&&
T(z):
\Vli\ot V_{s_N}\ot\cdots\ot V_{s_1}\ot\Vlj^{\ast a}
\ra
\Vlip\ot V_{s_N}\ot\cdots \ot V_{s_1}\ot\Vljp^{\ast a}
\nonumber
\\
&&
T(z)=
\Phi^\ast(z)
\check{R}_{1,s_N}(z)\cdots\check{R}_{1,s_1}(z)
\Phi(z).
\nonumber
\end{eqnarray}
The space $\Vli\ot V_{s_N}\ot\cdots\ot V_{s_1}\ot\Vlj^{\ast a}$
is identified
with $\H_{s_N\cdots s_1,ij}$ by the map
$\Phi^{s_N,\cdots,s_1}(1,\cdots,1)$ and its inverse.
Let us calculate the map $\tilde{T}(z)$ for which
\[
\begin{array}{ccc}
\H_{s_N\cdots s_1,ij}
&
\stackrel{\Phi^{s_N,\cdots,s_1}(1,\cdots,1)}{\ra}
&
\Vli\ot V_{s_N}\ot\cdots\ot V_{s_1}\ot\Vlj^{\ast a}
\\
\downarrow \tilde{T}(z)
&
{}
&
\downarrow T(z)
\\
\H_{s_N\cdots s_1,i+1j+1}
&
\stackrel{\Phi^{s_N,\cdots,s_1}(1,\cdots,1)}{\ra}
&
\Vlip\ot V_{s_N}\ot\cdots\ot V_{s_1}\ot\Vljp^{\ast a}
\end{array}
\]
is a commutative diagram.
Using the commutation relations \ref{ocom} we have
\begin{eqnarray}
\tilde{T}(z)&=&
\Phi^{s_N,\cdots,s_1}(1,\cdots,1)^{-1}
T(z)
\Phi^{s_N,\cdots,s_1}(1,\cdots,1)
\nonumber
\\
&=&
\Phi^{s_N,\cdots,s_1}(1,\cdots,1)^{-1}
\Phi^\ast(z)
\check{R}_{1,s_N}(z)\cdots\check{R}_{1,s_1}(z)
\Phi(z)
\Phi^{s_N,\cdots,s_1}(1,\cdots,1)
\nonumber
\\
&=&
A_{i,N}(z)
\Phi^\ast(z)
\Phi^{s_N,\cdots,s_1}(1,\cdots,1)^{-1}
\Phi^{s_N,\cdots,s_1}(1,\cdots,1)
\Phi(z)
\nonumber
\\
&=&
A_{i,N}(z)(1\ot T_{XXZ}(z)),
\nonumber
\end{eqnarray}
where
\begin{eqnarray}
&&
A_{i,N}(z)=
(-1)^{N+\sum_{j=1}^Ns_j}
q^{3\sum_{j=1}^N[{s_j-\overline{i+j-1}\over2}]-\sum_{j=1}^Ns_j+N
+[{N+i\over2}]-\sum_{j=1}^Ns_j\overline{i+j-1}}.
\nonumber
\end{eqnarray}
Hence, up to a scalar factor, the transfer matrix coinsides with
$1\ot T_{XXZ}(z)$. If we normalize the eigenvalue of the
vacuum vectors is equal to one, then the transfer matrix
is given by $1\ot T_{XXZ}(z)$.

\section{Discussion}
In this paper we introduce new kinds of q-vertex operators
and using them propose the mathematical model of the
inhomogeneous vertex models of the 6-vertex type.
One of our vertex operators $\phinnp(z)$ already appeared
in Miki's paper\cite{M} in the simplest non-trivial
form $n=1$ in a different context.

It follows from our mathematical setting of the models
that the excitation energies over the ground states
are the same as that of the 6-vertex model.
In our approach the impurity contributions to the several
physical quantities may be calculated through the
correlation functions.

As in the case of the other solvable lattice models\cite{FJMMN,JMN}
the trace of the compositions of the new vertex operators
satisfy certain q-difference equations.
Except the case of the form
$tr_{\Vli}(q^{-2\rho}\Phi(z_1)\cdots\Phi(z_k)\phidusms(z))$,
those equations are different from the q-KZ equation
with mixed spins.
Hence the situation is rather unexpected from the point of view
by the rough pictorial arguments\cite{JMN,FJMMN}.

The new vertex operators can be considered as non-local
operators actiong on the physical space of the XXZ-model.
This fact may open the door to study the fusion model\cite{RS,KNS}
of the 6vertex model using the vertex operators defined here.

Obviously we can introduce the inhomogeneities in the
spectral parameter (or the rapidities). This corresponds to
consider the space
$(V_{s_N-1})_{z_N}\ot\cdots\ot(V_{s_1-1})_{z_1}\ot
\Vli\ot\Vlj^{\ast a}$ etc.

\vskip2truecm
\noindent
{\bf Acknowledgement}\par
\noindent
I am grateful to N. Reshetikhin for stimulating discussions.
\vskip1truecm

\appendix
\section{Appendix1}
In this section we give the integral formula for the matrix
element $<\La_{i+1}|\ozu|\La_i>$.
\begin{eqnarray}
&&
<\La_{i+1}|\ozu|\La_i>
\nonumber
\\
&&
={1\over f}<\La_{i+1}|
\Phi_{\ep_1}(z_1)\cdots\Phi_{\ep_{n+1}}(z_n)
\Psi_{\mu_n}(u_n)\cdots\Psi_{\mu_1}(u_1)
|\La_i>
\nonumber
\\
&&
=
(-1)^{s_2}(q-q^{-1})^{r_1+s_2}
(-q)^{-i[{n\over2}]+(i-1)[{n+1\over2}]}
\prod_{j:\hbox{odd}}^{n+1}(-q^3z_j)^i
\prod_{j:\hbox{even}}^{n+1}(-q^3z_j)^{{1\over2}}
\prod_{j=1}^{n+1}(-q^3z_j)^{{s_2-s_1\over2}}
\nonumber
\\
&&
\prod_{k:\hbox{even}}^{n}(-qu_k)^{{1\over2}-i}
\prod_{a}(q^2z_a)^{-1}
\prod_b\prod_{j<b}(-q^3z_j)^{{1\over2}}
\prod_a\prod_{j<a}(-q^3z_j)^{-{1\over2}}
\prod_c\prod_{j>c}(-qu_j)^{{1\over2}}
\prod_d\prod_{j>d}(-qu_j)^{-{1\over2}}
\nonumber
\\
&&
\prod_a\int_{C_a}{dw_a\over2\pi i}
\prod_d\int_{C_d}{d\xi_d\over2\pi i}
\prod_aw_a^{-i+s_1-s_2}
\prod_{a<b}w_a^{-1}
\prod_{a<a'}w_a
\prod_{d}\xi_d^{i-1}
\prod_{d>c}\xi_d^{-1}
\prod_{d>d'}\xi_d
\nonumber
\\
&&
{
\prod_d\prod_{j=1}^{n+1}(1-{\xi_d\over q^3z_j})
\prod_a\prod_{l=1}^{n}(1-{qu_l\over w_a})
\over
\prod_a\prod_{j\leq a}(1-{w_a\over q^2z_j})
\prod_a\prod_{j\geq a}(1-{q^4z_j\over w_a})
\prod_d\prod_{k\leq d}(1-{u_k\over \xi_d})
\prod_d\prod_{k\geq d}(1-{\xi_d\over q^2u_k })
}
\nonumber
\\
&&
{
\prod_{a<a'}(1-{w_{a'}\over w_a})(1-{q^2w_{a'}\over w_a})
\prod_{d<d'}(1-{\xi_{d'}\over \xi_d})(1-{\xi_{d'}\over q^2\xi_d})
\over
\prod_{a,d}(1-{q\xi_d\over w_a})(1-{\xi_d\over qw_a}).
}
\end{eqnarray}
Here $r_1,r_2,s_1,s_2,a,b,c,d$ is defined as follows.
\begin{eqnarray}
&&
\{a\}=\{j|\ep_j=0\},
\quad
\{b\}=\{j|\ep_j=1\},
\quad
\{c\}=\{j|\mu_j=0\},
\quad
\{d\}=\{j|\mu_j=1\},
\nonumber
\\
&&
r_1=\sharp\{a\},
\quad
r_2=\sharp\{b\},
\quad
s_1=\sharp\{c\},
\quad
s_2=\sharp\{d\}.
\nonumber
\end{eqnarray}
$w_a$ and $\xi_d$ are the integral variables.
The integral contour $C_a$ and $C_d$ are taken in the following manner.
\begin{eqnarray}
C_a&:&\hbox{ $q^4z_j(j\geq a)$ and $q^{\pm1}\xi_d$(all $d$) are inside,}
\nonumber
\\
&:&
\hbox{ $q^2z_j(j\leq a)$ are outside.}
\nonumber
\\
C_d&:&\hbox{ $u_k(k\leq d)$ are inside,}
\nonumber
\\
&:&
\hbox{ $q^2u_k(k\geq d)$ and $q^{\pm1}w_a$(all $a$) are outside.}
\nonumber
\end{eqnarray}
%Note that the integrand does not have a residue at infinity
%in each variable $w_a$ and $\xi_d$.

The special components are given by
\begin{eqnarray}
&&
<\La_{1}|\ozu_{1\cdots1,0\cdots0}|\La_0>=
\nonumber
\\
&&
(-q)^{-[{n+1\over2}]}
\prod_{j:even}(-q^3z_j)^{{1\over2}}
\prod_{j=1}^{n+1}(-q^3z_j)^{{1-j\over2}}
\prod_{k:even}(-qu_k)^{{1\over2}}
\prod_{k=1}^n(-qu_k)^{{k-1\over2}},
\label{inti}
\\
&&
<\La_{0}|\ozu_{0\cdots0,1\cdots1}|\La_1>=
\nonumber
\\
&&
(-q)^{-[{n\over2}]+{n(n+1)\over2}}
\prod_{j:odd}(-q^3z_j)^{{1\over2}}
\prod_{j=1}^{n+1}(-q^3z_j)^{-{j\over2}}
\prod_{k:odd}(-qu_k)^{{1\over2}}
\prod_{k=1}^n(-qu_k)^{1-{k\over2}}.
\label{intii}
\end{eqnarray}

\def\gam{\gamma({z_1\over z_2})}
\def\bet{\beta({u_1\over u_2})}
\def\alp{\alpha({z\over u})}
\def\omg{\omega({u\over z})}

\section{Appendix2}
We give the description of the level one vertex operators
$\Phi(z)$ and $\Psi(z)$ on the free field realization of the
represenattions\cite{JMMN}.
\begin{eqnarray}
&&
\Phi_1(z)=
\exp\sum_{n=1}^\infty
\big({a_{-n}\over[2n]}q^{{7n\over2}}z^n\big)
\exp\sum_{n=1}^\infty
\big(-{a_n\over[2n]}q^{-{5n\over2}}z^{-n}\big)
e^{{\alpha\over2}}(-q^3z)^{\partial_{\alpha}+i\over2},
\nonumber
\\
&&
\Phi_0(z,w)=
{(q-q^{-1})(q^2z)^{-1}\over (1-{w\over q^2z})(1-{q^4z\over w})}
\exp\sum_{n=1}^\infty
\big({a_{-n}\over[2n]}q^{{7n\over2}}z^n
-{a_{-n}\over[n]}q^{{n\over2}}w^n
\big)
\nonumber
\\
&&\quad\quad\quad\quad\quad\quad
\exp\sum_{n=1}^\infty
\big(-{a_n\over[2n]}q^{-{5n\over2}}z^{-n}
+{a_{n}\over[n]}q^{{n\over2}}w^{-n}
\big)
e^{-{\alpha\over2}}w^{-\partial_{\alpha}}
(-q^3z)^{\partial_{\alpha}+i\over2},
\nonumber
\\
&&
\Psi_0(u)=
\exp\sum_{n=1}^\infty
\big(-{a_{-n}\over[2n]}q^{{n\over2}}u^n\big)
\exp\sum_{n=1}^\infty
\big({a_n\over[2n]}q^{-{3n\over2}}u^{-n}\big)
e^{-{\alpha\over2}}(-qu)^{-\partial_{\alpha}+i\over2}
(-q)^{-1+i},
\nonumber
\\
&&
\Psi_1(u,\xi)=
{-(q-q^{-1})\xi^{-1}\over (1-{u\over \xi})(1-{\xi\over q^2u})}
\exp\sum_{n=1}^\infty
\big(-{a_{-n}\over[2n]}q^{{n\over2}}u^n
+{a_{-n}\over[n]}q^{-{n\over2}}\xi^n
\big)
\nonumber
\\
&&\quad\quad\quad\quad\quad\quad
\exp\sum_{n=1}^\infty
\big({a_n\over[2n]}q^{-{3n\over2}}u^{-n}
-{a_n\over[n]}q^{-{n\over2}}\xi^{-n}
\big)
e^{{\alpha\over2}}\xi^{\partial_\alpha}
(-qu)^{-\partial_{\alpha}+i\over2}
(-q)^{-1+i},
\nonumber
\\
&&
\Phi_0(z)=
\int_{C_1}{dw\over 2\pi i}\Phi_0(z,w),
\nonumber
\\
&&
\Psi_1(u)=
\int_{C_2}{d\xi\over 2\pi i}\Psi_1(u,\xi),
\nonumber
\end{eqnarray}
where the contour $C_1$ and $C_2$ are specified by
\begin{eqnarray}
&&
C_1:\hbox{$q^4z$ is inside and $q^2z$ is outside},
\nonumber
\\
&&
C_2:\hbox{$u$ is inside and $q^2u$ is outside}.
\nonumber
\end{eqnarray}

\section{Appendix3}
Here we give the OPE of the level one vertex operators.
Notations are the same as that in \cite{JMMN} except that
the normal orderings are carried out for $e^{n\alpha}$
and $\partial_\alpha$.

\begin{eqnarray}
&&
\Phi_1(z_1)\Phi_1(z_2)
=\gam(-q^3z_1)^{{1\over2}}:\Phi_1(z_1)\Phi_1(z_2):,
\nonumber
\\
&&
\Phi_1(z_1)\Phi_0(z_2,w)
=\gam
{(-q^3z_1)^{-{1\over2}}\over1-{w\over q^2z_1}}
:\Phi_1(z_1)\Phi_1(z_2,w):,
\nonumber
\\
&&
\Phi_0(z_1,w)\Phi_1(z_2)
=\gam
{w^{-1}(-q^3z_1)^{{1\over2}}\over1-{q^4z_2\over w}}
:\Phi_0(z_1,w)\Phi_1(z_2):,
\nonumber
\\
&&
\Phi_0(z_1,w_1)\Phi_0(z_2,w_2)
=\gam w_1(-q^3z_1)^{-{1\over2}}
{(1-{w_2\over w_1})(1-{q^2w_2\over w_1})
\over
(1-{w_2\over q^2z_1})(1-{q^4z_2\over w_1})}
:\Phi_0(z_1,w_1)\Phi_0(z_2,w_2):,
\nonumber
\\
&&
\Psi_0(u_1)\Psi_0(u_2)=
\bet(-qu_1)^{{1\over2}}
:\Psi_0(u_1)\Psi_0(u_2):,
\nonumber
\\
&&
\Psi_0(u_1)\Psi_1(u_2,\xi)=
\bet
{(-qu_1)^{-{1\over2}}\over1-{\xi\over q^2u_1}}
:\Psi_0(u_1)\Psi_1(u_2,\xi):,
\nonumber
\\
&&
\Psi_1(u_1,\xi)\Psi_0(u_2)=
\bet
{\xi^{-1}(-qu_1)^{{1\over2}}\over1-{u_2\over \xi}}
:\Psi_1(u_1,\xi)\Psi_0(u_2):,
\nonumber
\\
&&
\Psi_1(u_1,\xi_1)\Psi_1(u_2,\xi_2)=
\bet \xi_1(-qu_1)^{-{1\over2}}
{(1-{\xi_2\over\xi_1})(1-{\xi_2\over q^2\xi_1})
\over
(1-{\xi_2\over q^2u_1})(1-{u_2\over \xi_1})}
:\Psi_1(u_1,\xi_1)\Psi_1(u_2,\xi_2):,
\nonumber
\\
&&
\Phi_1(z)\Psi_0(u)=
\alp
(-q^3z)^{-{1\over2}}
:\Phi_1(z)\Psi_0(u):,
\nonumber
\\
&&
\Phi_1(z)\Psi_1(u,\xi)=
\alp(-q^3z)^{{1\over2}}
(1-{\xi\over q^3z})
:\Phi_1(z)\Psi_1(u,\xi):,
\nonumber
\\
&&
\Phi_0(z,w)\Psi_0(u)=
\alp w(-q^3z)^{-{1\over2}}
(1-{qu\over w})
:\Phi_0(z,w)\Psi_0(u):,
\nonumber
\\
&&
\Phi_0(z,w)\Psi_1(u,\xi)=
\alp w^{-1}(-q^3z)^{{1\over2}}
{(1-{qu\over w})(1-{\xi\over q^3z})
\over
(1-{q\xi\over w})(1-{\xi\over qw})}
:\Phi_0(z,w)\Psi_1(u,\xi):,
\nonumber
\\
&&
\Psi_0(u)\Phi_1(z)=
\omg(-qu)^{-{1\over2}}
:\Psi_0(u)\Phi_1(z):,
\nonumber
\\
&&
\Psi_0(u)\Phi_0(z,w)=
\omg(-qu)^{{1\over2}}
(1-{w\over qu})
:\Psi_0(u)\Phi_0(z,w):,
\nonumber
\\
&&
\Psi_1(u,\xi)\Phi_1(z)=
\omg\xi(-qu)^{-{1\over2}}
(1-{q^3z\over \xi})
:\Psi_1(u,\xi)\Phi_1(z):,
\nonumber
\\
&&
\Psi_1(u,\xi)\Phi_0(z,w)=
\omg\xi^{-1}(-qu)^{{1\over2}}
{(1-{q^3z\over \xi})(1-{w\over qu})
\over
(1-{qw\over \xi})(1-{w\over q\xi})}
:\Psi_1(u,\xi)\Phi_0(z,w):.
\nonumber
\end{eqnarray}
Here
\begin{eqnarray}
&&
\gamma(z)={(q^2z^{-1})_\infty\over (q^4z^{-1})_\infty},
\quad
\beta(z)={(z^{-1})_\infty\over (q^2z^{-1})_\infty},
\quad
\alpha(z)={(qz^{-1})_\infty\over (q^{-1}z^{-1})_\infty},
\quad
\omega(z)={(q^5z^{-1})_\infty\over (q^3z^{-1})_\infty}.
\nonumber
\end{eqnarray}

\end{document}